%% file: template.tex
\documentclass[preprint]{vgtc}               % preprint
\ifpdf%                                % if we use pdflatex
  \pdfoutput=1\relax                   % create PDFs from pdfLaTeX
  \usepackage{graphicx}                % allow us to embed graphics files
  \DeclareGraphicsExtensions{.pdf,.png,.jpg,.jpeg} % for pdflatex we expect .pdf, .png, or .jpg files
\else%                                 % else we use pure latex
  \usepackage{graphicx}                % allow us to embed graphics files
  \DeclareGraphicsExtensions{.eps}     % for pure latex we expect eps files
\fi%

\graphicspath{{figures/}{pictures/}{images/}{./}} % where to search for the images

\usepackage{microtype}                 % use micro-typography (slightly more compact, better to read)
\PassOptionsToPackage{warn}{textcomp}  % to address font issues with \textrightarrow
\usepackage{textcomp}                  % use better special symbols
\usepackage{mathptmx}                  % use matching math font
\usepackage{times}                     % we use Times as the main font
\usepackage{cite}                      % needed to automatically sort the references
\usepackage{tabu}                      % only used for the table example
\usepackage{booktabs}                  % only used for the table example
\usepackage{amsmath}
\DeclareMathAlphabet{\mathcal}{OMS}{cmsy}{m}{n}
\usepackage{svg}
\usepackage{amssymb}
\usepackage[ruled,vlined]{algorithm2e}
\usepackage{url}
\usepackage{hyperref}
\usepackage{textcomp}
\usepackage{xcolor} % text color
\usepackage{braket}

\DeclareMathOperator*{\argmin}{arg\,min}

\onlineid{1315}

\vgtccategory{Research}

\vgtcinsertpkg

\title{ENI: Quantifying Environment Compatibility\\for Natural Walking in Virtual Reality}

\author{Niall L. Williams\thanks{e-mail: niallw@umd.edu}\\ %
        \scriptsize University of Maryland, College Park %
\and Aniket Bera\thanks{e-mail: bera@umd.edu}\\ %
     \scriptsize University of Maryland, College Park %
\and Dinesh Manocha\thanks{e-mail: dmanocha@umd.edu}\\ %
     \scriptsize University of Maryland, College Park}

\teaser{
  \centering
  \includegraphics[width=0.9\linewidth]{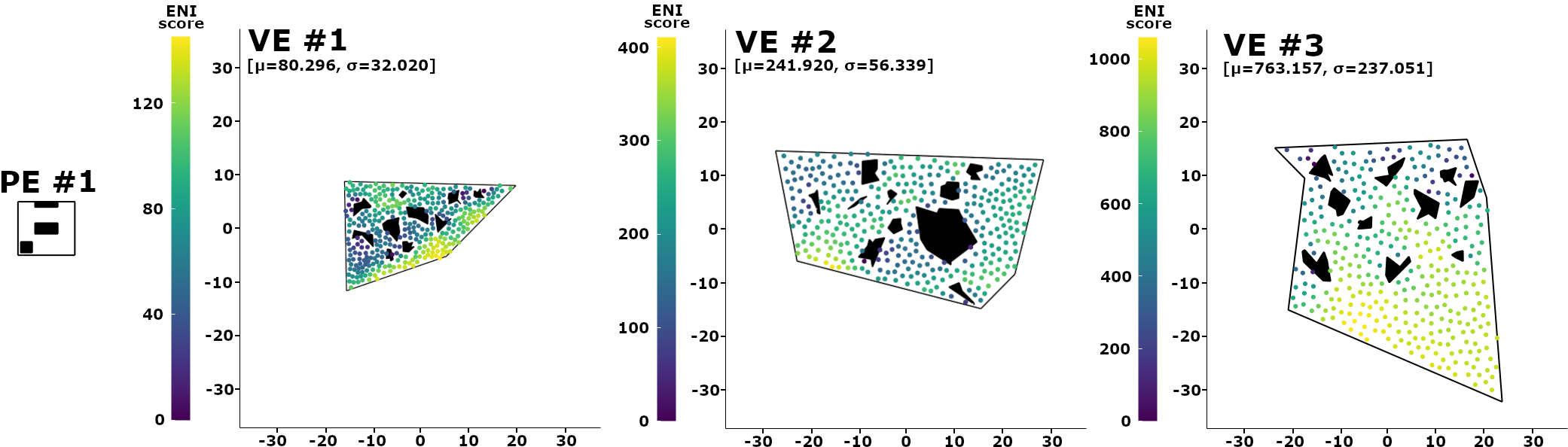}
  \caption{A visualization of our Environment Navigation Incompatibility (ENI) scores for physical environment \textcolor{black}{paired with three different virtual environments with increasing environment area.} 
  Our metric is used to accurately quantify whether it is possible to compute a good mapping between the geometric layouts of these environment.  
  We sample points across the virtual environment to represent the user's position (shown as colored circles) and compute the corresponding point in the physical environment based on comparing the local neighborhoods.
  \textcolor{black}{Overall, our metric helps us to determine which regions or subsets of the VE are more compatible with the PE.
  Through this visualization, we can see which regions of the VE are more or less compatible with the PE.}
  }
	\label{fig:teaser}
}

\abstract{We present a novel metric to analyze the similarity between the physical environment and the virtual environment for natural walking in virtual reality. Our approach is general and can be applied to any pair of physical and virtual environments. We use geometric techniques based on conforming constrained Delaunay triangulations and visibility polygons to compute the Environment Navigation Incompatibility (ENI) metric that can be used to measure the complexity of performing simultaneous navigation. We demonstrate applications of ENI for highlighting regions of incompatibility for a pair of environments, guiding the design of the virtual environments to make them more compatible with a fixed physical environment, and evaluating the performance of different redirected walking controllers. We validate the ENI metric using simulations and two user studies. Results of our simulations and user studies show that in the environment pair that our metric identified as more navigable, users were able to walk for longer before colliding with objects in the physical environment. Overall, ENI is the first general metric that can automatically identify regions of high and low compatibility in physical and virtual environments. Our project website is available at \href{https://gamma.umd.edu/eni/}{\textcolor{blue}{\texttt{https://gamma.umd.edu/eni/}}}.%
} % end of abstract

\CCScatlist{
  \CCScatTwelve{Navigability Metric}{Locomotion Interfaces}{Visibility Polygon}{Isovist};
}

\begin{document}


\maketitle

%% \section{Introduction} %for journal use above \firstsection{..} instead

\input{intro}
\input{background}

\input{methods}

\input{benefits}

\input{validation}
\input{conclusion}

% %% if specified like this the section will be committed in review mode
% TODO}

%\bibliographystyle{abbrv}
\bibliographystyle{abbrv-doi}

\bibliography{template}

\input{supplementary}

\end{document}

%% file: intro.tex
\section{Introduction}
\label{sec:introduction}
% \textbf{[include citations to classic VR natural walking papers that show that the important factor of walking in VR is the (1) relative size of the PE and VE and (2) the relative density/locations of obstacles in the environments]}

\textcolor{black}{Locomotion, the ability to explore a space, is a fundamental task in virtual reality (VR).}
Locomotion interfaces are techniques that enable a user to explore a virtual environment (VE).
\textcolor{black}{The main goal of locomotion interfaces is to} allow users to comfortably and safely explore the VE, which may be very large and dynamic, while they are located in a small physical environment (PE).
\textcolor{black}{While many locomotion interfaces have been developed \cite{di2021locomotion}, interfaces that allow users to explore VEs using natural walking are often preferred since they afford a higher sense of presence~\cite{usoh1999walking} and tend to lead to better performance at tasks in VR  applications~\cite{hodgson2008redirected, ruddle2009benefits}.
Although natural walking interfaces have many benefits, not all virtual tasks and pairings of physical and virtual environments are best suited for a natural walking interface \cite{di2021locomotion, suma2009evaluation}.
If the PE is prohibitively small or has a high number of obstacles, the user may have a more comfortable virtual experience with a locomotion interface that does not involve natural walking (such as teleportation or joystick movement).}

% \textbf{[dinesh suggests removing this]}
% This lack of a ``one-size-fits-all'' locomotion interface highlights the importance of evaluating the efficacy of locomotion interfaces for a particular virtual experience (or vice-versa, the efficacy of an interface in a variety of environments).
% The most straightforward method for such an evaluation is to conduct user studies to collect data on the user experience during locomotion.
% While this is an effective method for understanding how easily users are able to explore a pair of physical and virtual environments, user studies are often too time consuming for researchers to continually conduct them during the development process.
% Furthermore, user studies tell us which interfaces perform well in the tested environments, but they may not always tell us \textit{why} an interface performs well or poorly in a certain environment.

\textcolor{black}{A key issue with respect to locomotion interfaces that use natural walking is to determine whether a particular pair of physical and virtual environments (denoted  $\langle$PE, VE$\rangle$) is
amenable to collision-free locomotion.
%a particular locomotion experience is important for developing more effective locomotion interfaces as well as for designing $\langle$PE, VE$\rangle$ pairs that are more amenable to collision-free locomotion.
It is well-known in the environmental psychology community that the layout (i.e., the geometric structure) of an environment influences the shapes of the paths that a  users travel \cite{wiener2007isovist, simpson2017quantifying,golledge1999wayfinding}.
Indeed, similar studies have shown that the user's perception of a virtual environment's complexity and navigation also depends on its layout~\cite{dalton2010judgments, franz2005exploring}.
Although we have some understanding of the effects of environment layout on navigation in either PE or VR, a key issue with such locomotion interfaces is simultaneous exploration of a physical \textit{and} virtual environment.
Some prior work has studied the effects of environment layout on the feasibility of collision-free navigation with natural walking in the context of redirected walking \cite{razzaque2001redirected}. However, such studies are limited due to ambiguity in terms of how they define the layout of the environments, or lack of simultaneous consideration of the layouts of the PE and VE \textit{relative to each other}.}

\textbf{Main Results:} We address the problem of understanding the influence of environment layouts on the \textcolor{black}{VR locomotion} experience based on natural walking.
%would help to mitigate this problem.
%That is, if
Our goal is to accurately quantify to what degree the layouts of a given PE and VE influence a user's ability to avoid collisions during locomotion. 
We introduce an Environment Navigation Incompatibility (ENI) metric, which quantifies the difficulty of performing collision-free VR navigation in a given $\langle$PE,~VE$\rangle$ pair.
%physical and virtual environments.
ENI works by uniformly sampling locations across the PE and VE, and computing the most compatible physical location for each sampled location in the VE.
This compatibility computation is based on the visibility polygon \cite{de1997computational}, which characterizes the local structure of an environment around a location.
We formulate ENI on the visibility polygon due its ability to  characterize environment layout, and to capture local features of an environment, which are also used by humans to navigate through environments.
The final output of ENI is an $n$-dimensional vector of real numbers, where $n$ is the number of sampled locations in the VE.
We compute the mean and standard deviation of this $n$-dimensional vector and automatically create interactive visualizations to summarize the output of ENI and make it more interpretable.
% REWRITE THIS NEXT SENTENCE BY SAYING: WE COMPUTE THE MEAN AND STANDARD DEVIATION OF THE METRIC FOR WHAT?
%To make it easier to interpret this high-dimensional vector, we summarize it using the mean and standard deviation, as well as an interactive visualization of the ENI metric.
Using ENI, we can better understand how different regions of the VE and PE contribute to collision-free navigation.
ENI highlights regions of low and high compatibility between the PE and VE without requiring us to collect any locomotion data in the environment pair.
To summarize, our main contributions are:
\begin{itemize}
    \item A novel metric that quantifies the ease of collision-free navigation for a given pair of physical and virtual environments. 
    ENI is based only on the geometric layout of the environments, making it computable for any static $\langle$PE, VE$\rangle$ pair, assuming the layouts of the environments are known. %meaning it is computable for any pair of static physical and virtual environments as long as we know the geometric layout of the PE and VE. 
    ENI is the first general \textcolor{black}{VR} navigability metric that \textcolor{black}{simultaneously} considers the layouts of the PE and VE relative to each other.
    
    \item We highlight multiple benefits of ENI, including \textcolor{black}{analyses of how changes in the VE influence navigability, guidelines on how to design VEs to be more amenable to navigation for a fixed PE,} and evaluation of the performance of RDW controllers.
    
    \item Evaluation of ENI using extensive simulations and two user studies.
    We validate that ENI is capable of identifying $\langle$PE,~VE$\rangle$ pairs that are amenable to collision-free navigation \textit{without} the need for any locomotion data. 
    
\end{itemize}

%% file: background.tex
\section{Background and Prior Work}
\label{sec:background}

\subsection{Navigability Metrics}
\label{subsec:navigability_metrics}
In this work, we define the navigability of an environment as the average distance an agent can walk before colliding with an obstacle, in all directions across all positions in the environment.
Extending this to VR, where the user is simultaneously located in a PE and VE, the navigability of a $\langle$PE,~VE$\rangle$ pair is the average distance the user can walk before colliding with a \textit{physical} obstacle, in all directions across all positions in both environments.
% We assume that users will move in the VE such that they avoid collisions with virtual objects, so our definition of navigability in VR is only concerned with collisions with physical objects.
% For this work, we define the navigability of an environment as the ease with which an agent can travel along collision-free paths in that environment.
% \textbf{[TODO: DEFINE THIS FORMALLY: FOR ALL POSITIONS IN THE ENVIRONMENT, FOR ALL DIRECTIONS OF LOCOMOTION FROM EACH POINT, HOW LONG DOES IT TAKE TO COLLIDE WITH AN OBJECT]}
Our goal is to develop a metric that can quantify this notion of navigability for a $\langle$PE,~VE$\rangle$ pair.
\textcolor{black}{Navigability in VR} depends on many factors, including the layouts of the environments, the user's path through the VE, and the user's cognitive load during locomotion.
While all of these features are important to consider when assessing navigability, in this work we only study the effect of the environments' layouts.
% In the literature, researchers often use the phrase ``environment complexity'' when discussing navigability, since it is assumed that more complexity leads to more difficulty in collision-free navigation.
In particular, we use the term ``navigability'' to refer to the difficulty of collision-free navigation; it is also common for researchers to use the term ``complexity'' to refer to the same idea.
% and is also related to the notion of complexity of an environment (CAN YOU GIVE REFERENCES ABOUT ENVIRONMENT COMPLEXITY).
%but we note that prior work on navigability often uses the term ``complexity'' to refer to the same idea.

Quantifying navigability has been studied in related fields, including robot navigation and environmental psychology.
The factors that contribute the most to navigability depend on the domain in which navigability is being evaluated.
Thus, when discussing navigability metrics, it is important to consider the context in which the metric is being developed, since this context will influence which features a metric emphasizes.

\subsubsection{Environmental Psychology}
Researchers in environmental psychology have developed metrics to better understand how humans perceive the navigability of indoor spaces.
%In their field, the goal is usually to understand how humans perceive the navigability of a space so that buildings can be designed to more effectively create a comfortable and efficient navigation experience for the occupants.
Wiener et al.~\cite{wiener2007isovist} characterized environment complexity using geometric properties of isovists (also known as visibility polygons), which are the 2D planar region of space in an environment that can be seen from a given location.
They found a correlation between participants' perception of the complexity of the environments and some properties of isovists in these environments, such as isovist jaggedness and area.
Stamps~\cite{stamps2005isovists} explored the relationship between isovist properties and humans' perception of enclosure or permeability of urban environments.
Stamps~\cite{stamps2004mystery} also conducted a meta-analysis that found correlations between perceptions of enclosure of the environment and properties of a human's location, such as horizontal distance to the nearest obstacle.
To better understand the relationship between human navigation and layouts for the entire environment (as opposed to only local features that are captured with isovists), researchers have proposed space syntax measures~\cite{hillier1984hanson, hillier1976space}. % to quantify the layouts of environments.
Haq et al.~\cite{haq2003just} and Peponis~\cite{peponis1990finding} showed relationships between the global structure of environments and humans' navigation behavior through them.
Our approach is motivated by these prior methods and our metric is designed to quantify environment structure both on a local and global scale using isovists (visibility polygons) and random sampling, respectively.

\subsubsection{Locomotion in Virtual Environments}
\label{subsubsec:rdw}
\textcolor{black}{One of the most popular locomotion interface that enables real walking is redirected walking (RDW) \cite{razzaque2001redirected}.
Thus, in this section we focus mainly on prior work studying real walking in VR using RDW.}
There is some work on understanding how the shape of an environment influences the efficacy of the RDW steering algorithm.
Azmandian et al.~\cite{azmandian2015physical} studied the effect of the size and shape of the tracking space on the number of times that users have to orient away from physical obstacles.
Messinger et al.~\cite{messinger2019effects} studied the effect of the size of the tracking space and shape on the number of resets during RDW.
They considered square PEs of varying sizes in addition to PEs with different shapes including rectangular, trapezoidal, cross- and L-shaped.
Their results showed that users were able to avoid more collisions as the PE size grew. Moreover, non-convex PEs like the cross and L-shaped rooms lead to more collisions than the convex PEs.
Lee et al.~\cite{lee2020optimal} also studied how RDW algorithms perform as the shape and size of the PE changes.
In their work, they considered square PEs of varied sizes, as well as PEs in the shape of a square, trapezoid, cross, circle, T, and L, each with a roughly equal area.
Lee et al.~\cite{lee2020optimal} observed that larger PEs lead to fewer collisions and that non-convex PE shapes like cross, T, and L lead to more collisions than the convex PEs.
Williams et al.~\cite{williams2021arc} introduced the Complexity Ratio (CR) metric to quantify the navigability of a $\langle$PE,~VE$\rangle$ pair.
CR is defined as the ratio of the average distance to the nearest object in the PE and VE, averaged across many sampled points.
They showed that as the environments become more complex, users incur more resets.
Our Environment Navigation Incompatibility metric is more general than the methods described in this section, and provides a more accurate indication of the navigability of a given $\langle$PE,~VE$\rangle$ pair.

\subsection{Shape Similarity}
\label{subsec:shape_similarity}
There is considerable work on shape analysis of objects and environments in geometric computing.
The more similar a PE and VE are in terms of geometric shape, the more likely it is that a collision-free path in the VE corresponds to a collision-free path in the PE.
Therefore, any measure on the the similarity of a PE and VE provides a proxy to measuring the likelihood that a user can travel on collision-free paths in that $\langle$PE,~VE$\rangle$ pair.
%Since we seek to develop a metric to measure the similarity of a PE and VE, we draw inspiration from the shape similarity literature.

Many metrics have been proposed for shape similarity.
%Shape similarity is a well-studied problem, so there are a multitude of different metrics.
The Hausdorff distance pairs points from one shape to points on the other shape, and the distance measure is the longest distance between a pair of points~\cite{huttenlocher1993comparing}.
Another popular method for comparing polygons is the turning function, which parameterizes a polygon by the lengths of its edges and the interior angles between adjacent edges, which simplifies the problem to the comparison of 1D functions~\cite{arkin1991efficiently}.
Symmetric difference is a similarity measure that takes into account the area of overlap of the two polygons.
For two polygons $A$ and $B$, the symmetric difference metric is defined as \texttt{area}$((A-B)\cup (B-A))$~\cite{veltkamp2000shape}.
In our approach, we also use the metric \texttt{area}$(A-B)$ to measure the similarity of two visibility polygons that represent the user's position and orientation in an environment.

Shape similarity for 3D objects and environments has also been extensively studied~\cite{cardone2003survey,biasotti2016recent}.
Osada et al.~\cite{osada2002shape} introduced the notion of a \textit{shape function}, which is a function that characterizes the shape of an object when it is computed over a sufficiently dense set of random points on the object's surface.
Using such a function, Osada et al. computed histograms of shape functions for different objects and reduced the 3D shape similarity problem to comparing histograms. Our approach is also motivated by such techniques and we define appropriate shape functions
%In this paper, we take a similar approach by 
%defining an shape function 
to characterize the structure of an environment. Moreover, we compare pairs of physical and virtual shape function values to measure the similarity of the PE and VE layouts.
Shape correspondence is a problem that is closely related to shape similarity.
In the correspondence problem, we wish to compute a mapping of features (such as points on a surface) of one object to features on the other object~\cite{van2011survey}.
When computing corresponding features, the mapping is usually defined according to geometric properties, and the mapping can be constrained in different ways, such as being a one-to-one or one-to-many mapping.
In our approach, we take inspiration from  shape correspondence literature by finding the corresponding location in one environment that has the best ``local similarity'' to a given location from the other environment.

\textcolor{black}{Analogous to the geometric shape similarity problem, the robot motion planning community has developed metrics to quantify the layouts of environments with respect to collision-free navigation.
Anderson et al.~\cite{anderson2007proposed} proposed a metric that is a combination of the entropy and compressibility of the environment.
% Environments with high entropy have more options for different routes that the robot can take.
% The compressibility of the environment refers to the ease with which the distribution of obstacles in the environment can be described, wherein simpler environments have a higher compressibility score.
Similar to entropy, Crandall~\cite{crandall2003towards} used the branching factor and environment clutter as a measure of complexity for maze environments.
Shell et al.~\cite{shell2003human} borrowed concepts from space syntax~\cite{hillier1984hanson, hillier1976space} to define complexity in terms of the distance between adjacent convex regions of the environment.
El-Hussieny et al.~\cite{el2015robotic} focused on robots that explore unknown environments, and proposed an environment complexity metric that measures the difference between the expected and actual number of locations the robot needs to visit in order to map out the entire environment.}

%% file: methods.tex
\section{Environment Navigation Incompatibility Metric}
\label{sec:methods}
Our goal is to formulate a metric that quantifies a user's ability to navigate with collision-free paths in a given $\langle$PE,~VE$\rangle$ pair.
Since our driving application is VR locomotion with RDW, our metric is designed such that it accounts for the factors that are important for RDW. Currently, we only take into account the geometric layouts of the PE and VE.
%Although navigability in VR depends on many such factors, in this work we limit our metric's scope to only consider factors related to the geometry of the environments, 
%since researchers and developers usually have the most control over the environment structure and since environment geometry is relatively easy to model.
In VR locomotion, the user's ability to walk on collision-free paths depends primarily on their proximity to obstacles in both environments, which is defined by their relative position and orientation.
%Thus, our metric needs compare the user's proximity to obstacles in the PE and VE, while also being sensitive to the user's position and orientation.
If the layouts of the PE and VE are similar, the user's proximity to the obstacles would exhibit similar characteristics and thereby make it easier to navigate without collisions.
We consider a $\langle$PE,~VE$\rangle$ pair to be \textit{compatible} for RDW when they have a high degree of similarity.
%Thus, the Environment Navigation Incompatibility metric provides a measure of how different two environments are.
In this section, we detail the steps involved in formulating our ENI metric and analyze them.
%how these steps account for the factors that are salient for assessing navigability in VR (position, orientation, and proximity to obstacles).

% step 1: find a way to capture local features
% step 2: find a way to capture local features across the entire environment
% step 3: find a way to compare local features between two environments
%     within step 3, we need to account for position and rotation. position is handled by checking pairs of points, rotation by rotating the polygon for maximum area overlap
\subsection{Environment Representation}
\label{subsec:representing_geometry}
Since our metric is based only on the layouts of the PE and VE, our goal is to use a general representation of environment geometry.
While environments typically consist of 3D objects, we only consider the 2D projections of the environment onto the plane.
%it is important that we have a representation of the environment geometry that is amenable to computation.
% MAKE IT CLEAR THAT THE PE AND VE COULD CONSIST OF 3D OBJECTS, BUT YOU ONLY CONSIDER 2D PROJECTIONS ON A PLANE!
To prevent an environment from being infinitely large, we represent an environment as a closed polygon $\mathcal{P}$, and obstacles in the environment as holes of $\mathcal{P}$.
We use many standard geometric concepts to represent the environment geometry and a user's position and orientation in an environment.
Throughout this paper, a subscript of $phys$ or $virt$ on the symbols below is used to clarify if the symbol belongs to the physical or virtual environment, respectively.
\begin{itemize}
    \setlength\itemsep{.1em}
    \item $ \mathcal{P} $: A closed polygon, specified as an ordered set of vertices $\{v_1, v_2, ..., v_m\}$, where consecutive vertices are connected with an edge. If $\mathcal{P}$ has holes, they are specified using simple polygons, which are also represented as ordered sets of vertices connected by edges.
    \item $E$: An environment, specified as a closed polygon, potentially with holes representing obstacles in the environment.
    \item $ p $: A location in an environment, specified as a vector in $\mathbb{R}^2$.
    \item $ \theta $: A user's orientation in an environment, in the range $[0, 2\pi)$.
    \item $q$: A user's configuration (or state) in an environment. Their configuration consists of a position $p$ and an orientation $\theta$.
    \item $ \mathcal{C}_{obs} $: The set of all states that correspond to the user as having collided with an obstacle in $E$ (also known as the obstacle space).
    \item $ \mathcal{C}_{free} $: The set of all states that correspond to the user not collided with an obstacle in $E$ (also known as the free space). This set corresponds with the set of points in $E$ that are not inside any holes (obstacles) of $E$. This represents the regions of $E$ that the user can walk in.
    \item $ Free_{phys},\ Free_{virt} $: The free space in the physical or virtual environment, respectively.
    % \item $ xxx $: xx
\end{itemize}

\subsection{User Position and Orientation}
\label{subsec:accounting_for_pos_orientation}
Since visual perception plays a large role in driving a person's locomotion experience~\cite{patla1997understanding}, our goal is to define a metric using a geometric representation that is congruent with what users see during navigation in an environment.
As such, we use the visibility polygon as a representation of the local surroundings that a user sees at a single time instance during navigation.
For a given point $p$ in the plane, the visibility polygon $\mathcal{P}$ is the set of all points in the plane that are visible from $p$.
The point from which the visibility polygon is computed is also known as the kernel, $k$.
For an environment $E$ and any point $p \in \mathcal{C}_{free}$, we know that $\mathcal{P} \subset \mathcal{C}_{free}$ by the definitions of $\mathcal{P}$ and $\mathcal{C}_{free}$.
For an environment with a set of obstacles $\mathcal{O}$, we have a set $S$ of line segments that denotes the boundaries of all obstacles in the environment.
The visibility polygon can be computed in $O(s\log s)$ time~\cite{suri1986worst}, where $s = |S|$, which makes it a fairly efficient representation of the user's local surroundings. 
% YOU ARE USING THE SYMBOL $n$ MULTIPLE TIMES: NUMBER OF SAMPLES OR THE NUMBER OF VERTICES OF A POLYGON. TRY TO USE DIFFERENT SYMBOLS

While the visibility polygon allows us to represent the user's surroundings at a single moment during locomotion, \textcolor{black}{a single visibility polygon} does not account for the different positions and orientations a user could have across \textcolor{black}{an} entire path. % that the user travels on during locomotion.
%That is, the visibility polygon only describes the geometry around a single point; 
An environment has many different positions at which the user can be located, and the local surroundings could be different for every unique position.
In the following subsections, we explain how we extend the notion of visibility polygons to account for the many different positions and orientations a user can have during locomotion, thus providing a way to summarize the \textit{entire} environment using \textit{local} features which are most prominent during locomotion (i.e. visibility polygons).

\subsubsection{Positions}
\label{subsubsec:position}
In order to represent the user's position relative to local obstacles \textit{and} account for the many different possible positions the user can be located at, we uniformly sample points in $\mathcal{C}_{free}$ and compute a visibility polygon at each sampled point.
For an environment represented by a polygon $\mathcal{P}$ (potentially with holes), $\mathcal{C}_{free}$ is represented as the set of points in $\mathcal{P}$ that are not in any holes of $\mathcal{P}$.

\begin{figure}[t]
    \centering
    \includegraphics[width=.4\textwidth]{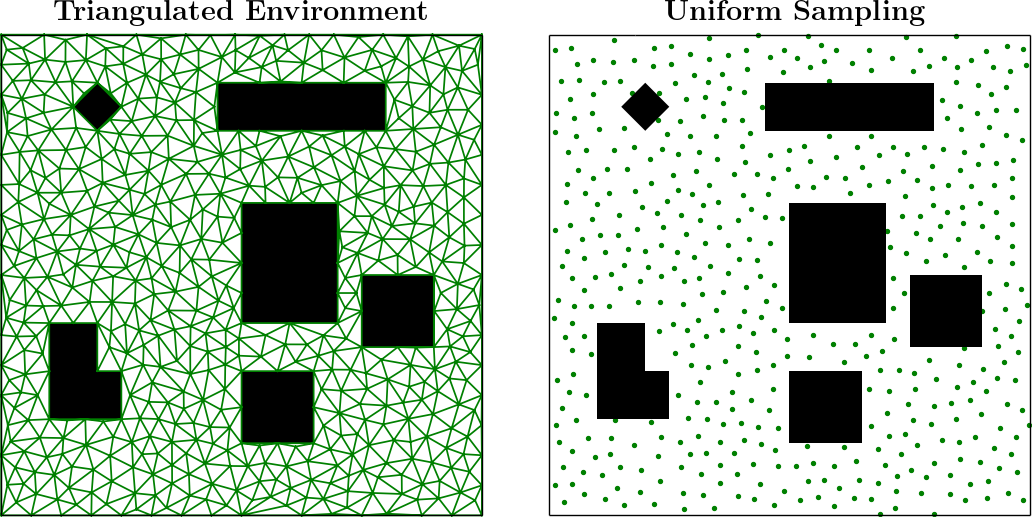}
    \caption{\textit{Left:} An environment with obstacles (black) and the constrained Delaunay triangulation (green) of the free space. \textit{Right:} The vertices (green) of the constrained Delaunay triangulation that lie inside the free space. These vertices are the sampled points at which we compute visibility polygons to describe the structure of the environment and compute our ENI metric.}
    % WHAT ARE YOU TRYING TO CONVEY WITH THIS FIGURE ABOUT YOUR METRIC COMPUTATION?  THE SIZE OF THE ENVIRONMENT IS IRRELEVANT. The environment size is $20m \times 20m$.}
    \label{fig:uniform_sampling}
\end{figure}

% DO YOU SAMPLE A GIVEN VISIBILITY POLYGON, OR YOU COMPUTE VISIBILITY POLYGON LOCATIONS FROM A LARGE NUMBER OF SAMPLES visibility polygons uniformly across $\mathcal{C}_{free}$.
% DO YOU COMPUTE A DELAUNAY TRIANGULATION OR A CONSTRAINED DELAUNAY TRIANGULATION. THE FORMER IS DEFINED FOR POINTS, WHILE THE LATER IS DEFINED FOR EDGES. HOW DO YOU REPRESENT C-FREE IN YOUR CASE?
First, we compute a conforming constrained Delaunay triangulation \cite{shewchuk1996triangle} of $\mathcal{C}_{free}$ with the constraint that each triangle has a maximum area, which is a free parameter that can be adjusted.\footnote{The implementation used is available here: \href{https://rufat.be/triangle/}{\texttt{rufat.be/triangle}}}
% of $0.1m$. MENTION AN AREA THRESHOLD. THE NUMBER 0.1m SHOULD BE IN SECTION 4 OR SO.
Once the triangulation is computed, we choose the sampled points as the set vertices of the triangulation that lie in the free space (see \autoref{fig:uniform_sampling}).
We denote this set of sampled points as $P = \{ p | p \in \mathcal{C}_{free} \text{ and } p \in \text{CDT} \}$, where CDT is the conforming constrained Delaunay triangulation of $E$.
Let $n$ be the number of points in $P$.
% We impose a maximum area constraint of $0.1m$ DO NOT MENTION THIS NUMBER HERE on the triangulation ensures that we obtain a reasonably dense sampling across the environment (the larger the area constraint, the sparser the sampling will be).
For each point $p \in P$, we compute a visibility polygon at $p$ (that is, the kernel $k$ of the visibility polygon is $p$).
% WHY DO YOU SAY "k = p, EXPLAIN YOUR CONSTRUTION TO JUSTIFY IT"
The final output of our uniform sampling is a set of $n$ visibility polygons  $\mathfrak{P} = \{ \mathcal{P}_1, \mathcal{P}_2, ..., \mathcal{P}_n \}$. 
% WHY DO YOU USE THE SYMBOL n HERE? WHAT DOES n CORRESPONDS TO?
In this manner, $\mathfrak{P}$ is an approximation of all the different local surroundings that the user can have in $E$ along any given path.

\textcolor{black}{We use a conforming constrained Delaunay triangulation to sample $\mathcal{C}_{free}$ since it tends to produce fairly uniformly-sized triangulations, which means the points we sample from the triangulation tend to be evenly spaced.
Other sampling methods like random sampling or importance sampling can be used, but care must be taken to ensure that the sampling scheme yields points that are evenly spread across the environment \textit{without} producing too many points, since this would slow down the metric computation process (see \autoref{subsec:metric_analysis}).
% may yield samplings with different resolutions in different regions of $\mathcal{C}_{free}$ or may be computationally impractical due to requiring too many samples to yield a uniform sampling.
In this work, we make no assumptions about which regions of $\mathcal{C}_{free}$ the user may be located in, so a uniform sampling is best suited for our metric computation.}
%as they their change location.

\setlength{\textfloatsep}{12.5pt plus 1.0pt minus 2.0pt}

\subsubsection{Orientations}
\label{subsubsec:orientation}
While the visibility polygons in $\mathfrak{P}$ represent the different local surroundings the user can perceive as they change their position in $E$, these polygons do not account for the fact that the user's perception of their surroundings will also depend on their orientation in the environment.
To account for the user's orientation at a position in $E$, we rotate the visibility polygon around its kernel (\textcolor{black}{since} the kernel also corresponds to the user's position in $E$).
For a visibility polygon $\mathcal{P}$ with kernel $k$ and vertices $\{ v_1, v_2, ..., v_m \}$, we define the visibility polygon rotated counterclockwise by $\theta$ radians as:
\begin{equation}
    \mathcal{P}^\theta = 
    \Set{ \left(
            \begin{bmatrix}
            \cos \theta & -\sin \theta \\
            \sin \theta & \cos \theta
            \end{bmatrix} 
            \begin{bmatrix}
            v.x - k.x \\
            v.y - k.y
            \end{bmatrix} 
        \right) + 
        \begin{bmatrix}
        k.x \\
        k.y
        \end{bmatrix} | v \in \mathcal{P} }
\end{equation}
Here, $v.x$ and $v.y$ represent the $x$- and $y$-coordinates of the vertex $v$, respectively. 
Thus, $\mathcal{P}^\theta$ has the same shape as that of $\mathcal{P}$, with the only difference being the orientation of this polygon in the plane.

\subsection{Measuring Compatibility of Local Surroundings}
\label{subsec:measuring_env_similarity}
\begin{figure}[t]
    \centering
    \includegraphics[width=.4\textwidth]{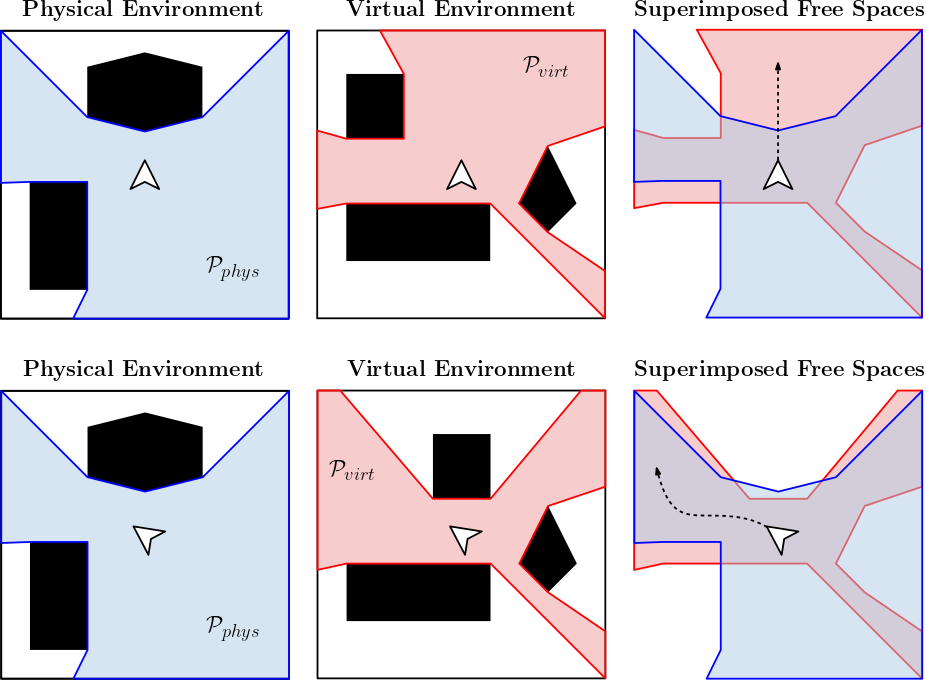}
    \caption{An illustration of the impact of the similarity between the user's physical and virtual surroundings on their ability to travel on collision-free paths. In the top row, the user (shown as the white cursor) cannot walk forward in the VE without colliding with an object in the PE. In the bottom row, the user's proximity to obstacles in the two environments is more similar, so more of the possible paths in the VE correspond to collision-free paths in the PE. In our metric, we compute this area of the virtual surroundings that cannot be accessed from a particular physical surrounding as a measure of the navigability at a pair of physical and virtual configurations.
    % EXPLAINED HOW SUPER-IMPOSED FREE SPACES USED HERE? WHAT CONCLUSIONS DO YOU MAKE FROM THOSE SPACES FOR YOUR METRIC? MAKE A CONNECTION TO YOUR METRIC 
   % Figure adapted from \cite{williams2021redirected}.
    }
    \label{fig:superimpose}
\end{figure}

% DO YOU WANT TO USE THE WORD COMPATIBILITY
% \textbf{[dinesh says cut down this section because its repetitive]}
% Recall that the ability for a user to walk on a collision-free path in VR depends partially on the differences in the layouts of the PE and VE.
% %That is, the more similar two environments are, the more likely it is that a collision-free path in one environment corresponds to a collision-free path in the other.
% This property extends to visibility polygons, which are a subset of an environment.
% Given that the visibility polygon is just a local representation of the environment (i.e. a subset of a path in an environment), we can make the same argument the impact of the similarity of visibility polygons on the user's ability to travel on a collision-free path.
The more similar the physical and virtual visibility polygons are, the more likely it is that the user is able to walk on collision-free paths in the local neighborhood of the current location (see \autoref{fig:superimpose}).
% This principle is illustrated in \autoref{fig:superimpose}.
Thus, to measure the compatibility of a user's physical and virtual surroundings, we need a way to measure the similarity of two visibility polygons.
This provides a way to assess the navigability of a user's configuration in the PE and VE.

As mentioned in \autoref{subsec:shape_similarity}, there are many different ways to measure the similarity of shapes.
% The particular definition of the shape similarity metric will depend on the problem domain \cite{van2011survey}.
In this work, we are mainly concerned with the user's proximity to obstacles.
Since the visibility polygon already encodes the proximity to obstacles in all directions, our notion of shape similarity  depends on the sizes of the polygons.
Thus, to measure the similarity (compatibility) of two visibility polygons, we measure the amount of overlapping area between the polygons.

Given a physical position $p_{phys} \in Free_{phys}$ and a virtual position $p_{virt} \in Free_{virt}$, let $\mathcal{P}_{phys}$ and $\mathcal{P}_{virt}$ be the visibility polygons with kernels $k_{phys} = p_{phys}$ and $k_{virt} = p_{virt}$, respectively.
We define the similarity of $\mathcal{P}_{phys}$ and $\mathcal{P}_{virt}$ as the area of $\mathcal{P}_{virt}$ that is ``inaccessible'' from $\mathcal{P}_{phys}$.
That is, if the user is standing at $k_{phys}$ and $k_{virt}$, our similarity metric for two visibility polygons is the total area of $\mathcal{P}_{virt}$ that cannot be reached due to occlusion by an edge of the boundary of $\mathcal{P}_{phys}$ (see \autoref{fig:superimpose}).
Formally, the similarity of visibility polygons $\mathcal{P}_1$ and $\mathcal{P}_2$ is computed using the Boolean difference operation~\cite{de1997computational}:
\begin{equation}
    \label{eqn:vis_poly_similarity_metric}
    \phi \left( \mathcal{P}_1, \mathcal{P}_2 \right) = \texttt{area} \left( \mathcal{P}_1 \setminus \mathcal{P}_2 \right).
\end{equation}
Here, the \texttt{area(}$x$\texttt{)} function returns the total area of the set of polygons $x$.
% WHAT IS X: A SINGLE POLYGON OR MULTIPLE POLYGONS?
Note that $\mathcal{P}_1$ and $\mathcal{P}_2$ must be translated such that their kernels lie at the same coordinates in the plane (i.e. $\mathcal{P}_1$ and $\mathcal{P}_2$ must be ``overlayed'' ontop of each other).
\textcolor{black}{A visualization of this similarity metric can be found in the supplementary materials (\autoref{fig:boolean_difference}).}
% This similarity metric is visualized in \autoref{fig:boolean_difference}.
If $\phi (\mathcal{P}_1, \mathcal{P}_2 ) = 0$, this implies that $\mathcal{P}_1$ is entirely contained inside $\mathcal{P}_2$, and that the user can reach all regions of $\mathcal{P}_1$ without colliding with any obstacles represented by the edges of $\mathcal{P}_2$.
The larger $\phi (\mathcal{P}_1, \mathcal{P}_2 )$ is, the more dissimilar $\mathcal{P}_1$ and $\mathcal{P}_2$ are, which increases the amount of space in $\mathcal{P}_1$ that is inaccessible when the user is standing at $k_2 \in \mathcal{P}_2$.
Note that $\phi$ is \textit{not} symmetric, so $\phi (\mathcal{P}_1, \mathcal{P}_2 ) \neq \phi (\mathcal{P}_2, \mathcal{P}_1 )$.
In our formulation, we only compute $\phi ( \mathcal{P}_{virt}, \mathcal{P}_{phys} )$, since our primary concern is regions of the VE that are inaccessible due to constraints imposed by the PE.
It should be noted that area is only one measure of the similarity of $\mathcal{P}_{phys}$ and $\mathcal{P}_{virt}$.
It is also possible to use other properties of $\mathcal{P}$ as the basis of our comparison, such as the average distance between the boundary of $\mathcal{P}$ and $k$, or the length of the shortest line connecting two points on the boundary of $\mathcal{P}$ that also passes through $k$.
In our benchmarks, we observed that area was an overall better metric because it provides a holistic summary of the differences between $\mathcal{P}_1$ and $\mathcal{P}_2$, while other metrics tended to ignore regions of either polygon.

% \subsection{Putting It All Together}
\subsection{ENI  Metric}
\label{subsec:putting_it_all_together}
% steps:
%     1) sample points across envs
%     2) for each sampled point in virt env, find best phys point
%     3) compute histogram/distribution of "best similarity" measure
% \textbf{[todo: add some kind of pseudo code or steps for the whole process, then divide this section according to those]}
%We now have all the components needed to compute our navigability metric: a robust representation of environment geometry (\autoref{subsec:representing_geometry}), a way to represent a user's local surroundings with respect to their position and orientation (\autoref{subsec:accounting_for_pos_orientation}),
% a way to extend this representation to account for variability in a user's position (\autoref{subsubsec:position}) and orientation (\autoref{subsubsec:orientation}), 
%and a similarity metric to measure how compatible the user's physical and virtual surroundings are for RDW (\autoref{subsec:measuring_env_similarity}).
In this section, we describe how various components described above are used to compute our Environment Navigation Incompatibility (ENI) metric. 
% An overview of this process is shown in \autoref{}.
Our goal is to estimate, for any possible virtual state $q_{virt}~\in~Free_{virt}$, which physical state $q_{phys}~\in~Free_{phys}$ is most compatible with $q_{virt}$.
If we can compute this for all states in $Free_{virt}$, we will have a measure that tells us how easy, in the ideal case, it will be for a user to navigate on a collision-free path in the given $\langle$PE,~VE$\rangle$ pair.
Our metric requires as input the 2D layouts of a physical environment $E_{phys}$ and a virtual environment $E_{virt}$ (\autoref{subsec:representing_geometry}).
Given the layouts of $E_{phys}$ and $E_{virt}$, we sample points uniformly across each environment using the technique detailed in \autoref{subsubsec:position}, with a maximum area constraint of $0.1m$.
For each sampled point $p \in E$, we compute the associated visibility polygon with kernel $p$.

This yields two sets of visibility polygons, $\mathfrak{P}_{phys}$ and $\mathfrak{P}_{virt}$, which are an approximation of all the possible states that the user can have in $E_{phys}$ or $E_{virt}$.
For each state $q_{virt} \in Free_{virt}$ (i.e. each visibility polygon $\mathcal{P}_{virt} \in \mathfrak{P}_{virt}$), we wish to find the physical state $q_{phys} \in Free_{phys}$ that is most similar to $q_{virt}$.
To do this, we compare each $\mathcal{P}_{phys} \in \mathfrak{P}_{phys}$ to each $\mathcal{P}_{virt} \in \mathfrak{P}_{virt}$ using the similarity metric in \autoref{eqn:vis_poly_similarity_metric}.
However, it is not enough to compute $\phi (\mathcal{P}_{virt}, \mathcal{P}_{phys})$ with the polygons $\mathcal{P}_{phys}$ and $\mathcal{P}_{virt}$.
The polygons computed from our uniform sampling represent the different positions the user can have, but they do not account for the user's orientation.
To measure the similarity of $\mathcal{P}_{phys}$ and $\mathcal{P}_{virt}$ while also accounting for the different orientations the user can have, we aim to solve the following optimization problem:
\begin{equation}
    \Phi^* \left( \mathcal{P}_{virt}, \mathcal{P}_{phys} \right) = \min_{\theta \in [0, 2\pi)} 
        \phi \left( 
            \mathcal{P}_{virt}, \mathcal{P}_{phys}^\theta
        \right).
\end{equation}
That is, we want to find the $\theta \in [0, 2\pi)$ that minimizes the value of $\phi (\mathcal{P}_{virt}, \mathcal{P}_{phys}^\theta)$ (i.e. maximizes the similarity between $\mathcal{P}_{virt}$ and $\mathcal{P}_{phys}$).
In practice, we found that computing $\Phi^* \left( \mathcal{P}_{virt}, \mathcal{P}_{phys} \right)$ for every pair of visibility polygons in $\{\mathfrak{P}_{virt} \times \mathfrak{P}_{phys}\}$ was too expensive.
To lower the computation time, we limit $\Phi^*$ to optimize $\phi$ in the domain $\Theta_\Delta~=~\{ \theta_1, \theta_2, ... \theta_{10} \}$, where $\theta_1 = 0^\circ$ and $\theta$ increases in increments of $36^\circ$.
Thus, for a pair of physical and virtual visibility polygons $\mathcal{P}_{phys}$ and $\mathcal{P}_{virt}$, we approximate the maximum similarity of the polygons as:
\begin{equation}
\label{eqn:best_compatibility}
    \Phi \left( \mathcal{P}_{virt}, \mathcal{P}_{phys} \right) = \min_{\theta \in \Theta_\Delta} 
        \phi \left( 
            \mathcal{P}_{virt}, \mathcal{P}_{phys}^\theta
        \right).
\end{equation}

Now we have everything necessary to approximate the optimal $q_{phys}$ for a given $q_{virt}$.
Using visibility polygons to represent $q_{phys}$ and $q_{virt}$, we compute the $q_{phys}$ that is most compatible with $q_{virt}$ as:
\begin{equation}
    \label{eqn:final_optimization}
    \mathcal{P}^*_{phys} = \argmin_{\mathcal{P}_{phys} \in \mathfrak{P}_{phys}}\Phi \left( \mathcal{P}_{virt}, \mathcal{P}_{phys} \right).
\end{equation}

% Given $\mathcal{P}^*_{phys}$ which best matches $\mathcal{P}_{virt}$, we finally compute the similarity metric between them as $\Phi(\mathcal{P}_{virt}, \mathcal{P}^*_{phys})$.
% By computing $\Phi(\mathcal{P}_{virt}, \mathcal{P}^*_{phys})$ for each $\mathcal{P}_{virt} \in \mathfrak{P}_{virt}$, we get an $n$-dimensional vector of scalar compatibility scores, where $n = |\mathfrak{P}_{virt}|$.
% This vector is the final output of the ENI metric.
% Since $n$ can be in the thousands, we summarize the output vector using the mean and standard deviation of the vector, which we denote as $[\mu, \sigma]$.
% Not that this summary does not perfectly characterize the ENI measure, since two measures can have the same mean and standard deviation.
% Methods of properly interpreting the ENI metric are discussed in \autoref{subsec:visualization}.
% This allows to fully define the ENI metric.
To compute the ENI metric, we compute the compatibility score between each virtual visibility polygon and its most compatible physical visibility polygon, yielding a vector of real numbers representing the best-case compatibility for each sampled state in $Free_{virt}$.
Formally, this is defined as:
% \begin{equation}
%     \label{eqn:metric}
%     \mathbf{x} = \left\{ \Phi \left( \mathcal{P}_{virt}, \mathcal{P}^*_{phys} \right) \middle| \mathcal{P}_{virt} \in \mathfrak{P}_{virt} \right\}
% \end{equation}
\begin{equation}
    \label{eqn:metric_computation}
    \mathbf{x} = \left\{ \Phi \left( \mathcal{P}_{virt}, \argmin_{\mathcal{P}_{phys} \in \mathfrak{P}_{phys}}\Phi \left( \mathcal{P}_{virt}, \mathcal{P}_{phys} \right) \right) \middle| \mathcal{P}_{virt} \in \mathfrak{P}_{virt} \right\},
\end{equation}
where $\mathbf{x}$ is an $n$-dimensional vector and $n=|\mathfrak{P}_{virt}|$.
This vector $\mathbf{x}$ is the final output of the ENI metric.
Since $n$ can be in the thousands, we summarize the output of ENI using the mean and standard deviation of the vector, which we denote as $[\mu, \sigma]$.
Note that this summary does not perfectly characterize the ENI measure, since two \textcolor{black}{distinct} measures can have the same mean and standard deviation.
Details on how to accurately interpret $\mathbf{x}$ are discussed in \autoref{subsec:visualization}.

\subsection{ENI Metric: Analysis}
\label{subsec:metric_analysis}
In this section, we discuss the properties of our Environment Navigation Incompatibility metric.
These properties help ensure that ENI avoids ambiguity and accurately models the important features of the VR locomotion problem during computation.

\textbf{Sensitivity to input}: Assuming the point sampling parameters are fixed (i.e. for a fixed input, computing the sampled points multiple times yields the same set of points each time), the output of the ENI metric is always the same.
This is because our metric performs an exhaustive search of all pairs of physical and virtual states $\{\mathfrak{P}_{phys}~\times~\mathfrak{P}_{virt}\}$ when computing the compatibility of the PE and VE.
As a result of this property, we avoid ambiguity that can arise from using environment properties that do not completely characterize the layout of the environment (e.g. environment area).
Note that since we compute a triangulation of each environment, small perturbations in the input geometry will result in different triangulations (and sampled points), which will yield slightly different metric measurements.

\textbf{Coupled computation:} The ENI metric requires a $\langle$PE,~VE$\rangle$ \textit{pair} as input in order to be computable.
This is because the metric is designed to compute the compatibility of the two environments.
ENI was intentionally designed in this way since it is the differences between the PE and VE that make collision-free navigation difficult.
This property ensures that our metric appropriately considers the layouts of the PE and VE \textit{relative to each other}, which makes ENI a more faithful measure of navigability in \textcolor{black}{VR}.

\textbf{Sampling density:} The ENI metric has one free parameter, which is the maximum area of triangles in the constrained Delaunay triangulation of $\mathcal{C}_{free}$ that is used to uniformly sample points in $E$ (\autoref{subsubsec:position}).
The smaller this parameter is, the denser the sampling of points in $E$.
A denser sampling yields a more more accurate measure for the ENI metric, but also increases the size of the output $\mathbf{x}$ and the computation time.
In our implementation, we set the maximum area such that each environment has roughly 500 samples (\textit{i.e.}, the maximum area parameter depends on the area of $\mathcal{C}_{free}$).
%However, we do not claim that this is the best value for the triangle area parameter.
% We chose a maximum area of $0.1m$ since it generated samples that were fairly dense in our benchmarks. % environments we tested.
% Furthermore, a circle with an area of $0.1m$ has a diameter of roughly $0.36m$, which is a close approximation of the average shoulder width of women over 20 years old~\cite{mcdowell2009anthropometric}, meaning our triangulation provides a sample that accurately approximates the gamut of different positions an adult user can have in an environment.

\textcolor{black}{To validate that 500 samples was sufficient, we computed the ENI measure for a $\langle$PE,~VE$\rangle$ pair with varying amounts of sample density, and compared the changes in the mean and standard deviation of the ENI measures.
The results are shown in \autoref{tab:max_area}.
From these results, we can see that increasing the number of points more than tenfold does not yield a noticeable change in the $\mu$ or $\sigma$ of the ENI metric, suggesting that our 500 samples points is sufficiently high resolution.
Furthermore, increasing the sample density leads to prohibitively high computation times for relatively little increase in metric accuracy, since the runtime complexity of the ENI metric is $O(nm)$, where $n=|\mathfrak{P}_{virt}|$ and $m=|\mathfrak{P}_{phys}|$.}

\setlength{\tabcolsep}{5pt}
\begin{table}[ht]
    \centering
    % \begin{tabular}{lcccc}
    %     $\langle$PE,~VE$\rangle$ pair & Sampled points & ENI $\mu$ & ENI $\sigma$ & Computation time (\textit{s})\\ \midrule
    %     (PE~\#1, VE \#1) & 257 & \textbf{todo} & \textbf{todo} & \textbf{todo} \\ 
    %     (PE~\#1, VE \#1) & 517 & \textbf{todo} & \textbf{todo} & \textbf{todo} \\ 
    %     (PE~\#1, VE \#1) & 1050 & \textbf{todo} & \textbf{todo} & \textbf{todo} \\
    %     (PE~\#1, VE \#1) & 2129 & \textbf{todo} & \textbf{todo} & \textbf{todo} \\
    %     (PE~\#1, VE \#1) & 4337 & \textbf{todo} & \textbf{todo} & \textbf{todo} \\ 
    %     \midrule \midrule
    % \end{tabular}
    \begin{tabular}{llll}
        Sampled points & ENI $\mu$ & ENI $\sigma$ & Computation time (\textit{s})\\ \midrule
        \textcolor{black}{257} & \textcolor{black}{486.9} & \textcolor{black}{92.3} & \textcolor{black}{168} \\ 
        \textcolor{black}{517} & \textcolor{black}{486.7} & \textcolor{black}{93.4} & \textcolor{black}{347} \\ 
        \textcolor{black}{1050} & \textcolor{black}{482.3} & \textcolor{black}{98.5} & \textcolor{black}{664} \\
        \textcolor{black}{2129} & \textcolor{black}{482.3} & \textcolor{black}{100.0} & \textcolor{black}{1277} \\
        \textcolor{black}{4337} & \textcolor{black}{481.8} & \textcolor{black}{100.1} & \textcolor{black}{2593} \\ 
        \midrule \midrule
    \end{tabular}
    \caption{\textcolor{black}{Effect of sample density on the accuracy of the ENI metric, using the $\langle$PE~\#1, VE \#1$\rangle$ environment pair. After increasing the density of our point sampling by roughly $16\times$, the mean and standard deviation of the ENI metric exhibited very little change in values, but suffered a significantly greater computation time.}}
    \label{tab:max_area}
\end{table}
%While we believe a maximum area of $0.1m$ is a suitable choice for RDW, this is a parameter that can be freely changed if one desires a coarser or finer sampling.
% \textbf{[TODO: TALK ABOUT THE EFFECT OF CHANGING THE SAMPLING DENSITY]} talk about n being low vs high, and mention that i dont know the best value for n. talk about the pros and cons of big/small n. give heuristic for how i picked the current n

% \subsection{Sensitivity to Input}
% % - the output depends entirely on the input

% \subsection{Coupled Computation}
% % - we consider both envs simultaneously

% \begin{enumerate}
%     \item 
% \end{enumerate}
% - deterministic: a given output will have only one output (subject to the DT area parameter).
% - based only on geometry
% - considers both envs simultaneously

%% file: benefits.tex
\section{Applications and Benefits of ENI}
\label{sec:benefits}

% The benefits of our metric include:
%     - allows us to estimate goodness of RDW without running lots of user studies.
%     - enables deeper understanding of the impact of env. layouts on collision-free navigation
%     - allows researchers to create virtual environments that are better suited for a given physical environment. VE optimization.
%     - allows for a new performance metric for RDW controllers: compute the average "compatibility metric" score across the paths that a given controller steers a user on
%         - note that this also opens the door for developing potentially improved RDW controllers
% Our environment compatibility metric provides a notion of the compatibility of a PE/VE pair with respect to RDW.
% This compatibility score helps us estimate to what degree a user will be able to travel along collision-free paths when located in the physical and virtual environments.
% In this section, we use our redirection metric to improve a user's experience during navigation in a virtual world.
%how our environment compatibility metric can be applied to gain insights that can improve the user experience during navigation with RDW.

% In this section we highlight some use-cases for our ENI metric.

\subsection{Analyzing Areas with Low and High Compatibility}
\label{subsec:visualization}
% DO YOU ASSUME A FIXED PHYSICAL WORLD, AND EVALUATE DIFFERENT VIRTUAL LAYOUTS! MAKE THAT CLEAR.
% - introduce our visualization tool
% - We can get a quick overview of the compatibility of the env pairs
% - we can dive deeper into specific locations of the environments using lasso tool
Since the output of ENI is an $n$-dimensional vector of real numbers, it is difficult to directly interpret the ENI measure.
To aid in interpretation, we visualize the ENI measure using an interactive visualization which can be seen in \textcolor{black}{the supplementary materials (\autoref{fig:vis_histogram}).}
% CAN YOU MENTION THE SPECIFIC TECHNIQUE USED TO VISUALIZE THIS d-DIMENSIONAL VECTOR ON A 2D PLANE,  DID YOU EXPLAIN THE COLOR SCHEME IN THAT FIGURE.
Our visualization includes a map of the PE and VE, and a histogram of the individual compatibility scores computed for each pair of physical and virtual visibility polygons (see \autoref{subsec:putting_it_all_together}). 
% HOW ARE THESE VISIBILITY POLYGONS COMPUTED IN THESE ENVIRONMENTS (POINT TO THE RELEVANT PARTS IN SECTION 3).
Each circle drawn in $E_{virt}$ represents the kernel of a virtual visibility polygon. 
% BUT THAT DEPENDS ON YOUR SAMPLING SCHEME.
A circle's color is determined by the compatibility score $\Phi$ computed from \autoref{eqn:best_compatibility} using the visibility polygon centered at that circle's location.
By taking a dense\textcolor{black}{, uniform} sampling of points and coloring them according to their compatibility scores, our visualization provides an easy way to see which regions of the VE lead to the most incompatibility with respect to the given PE.

To further improve the interpretability of the ENI metric, our visualization also includes interactive tools that allow researchers to explore the metric output.
% WHAT DO YOU MEAN? IS VERY FINE VISUALIZATION BRINGS ANYTHING NEW HERE? 
With a lasso tool (\autoref{fig:viz_lasso} in supplementary materials), users can select points in $E_{virt}$ which will also highlight the corresponding most compatible points in $E_{phys}$ that were computed from \autoref{eqn:final_optimization}.
% (see \autoref{fig:viz_lasso}).
This helps researchers to understand, for a selected group of configurations in the VE, which configurations in the PE will be most amenable to collision-free navigation.
% HOW DO YOU USE THOSE RESULTS.
Additionally, hovering over any bar in the histogram will highlight the physical and virtual points that contributed to the selected histogram bar.
% (see \autoref{fig:vis_histogram}).
Using this interactive visualization, researchers can explore the compatibility of a pair of environments on both a broad and a specific level, which makes it easier to design improved RDW controllers and more compatible $\langle$PE,~VE$\rangle$ pairs.

\subsection{\textcolor{black}{Analysis of Changes in VE on Compatibility}}
\label{subsec:VE_changes}
\textcolor{black}{
The ENI metric helps us to understand how different changes to the VE can effect the ease of collision-free navigation in the given PE.
We assume that the PE is fixed, since this is usually true in practice.
To demonstrate how ENI can be used to understand the effects of VE changes on navigability, we show the results of two examples.
First, we look at the effects of changing the density of objects in the VE on navigability, and second we look at the effects of changing the size of the VE on navigability.
}

\subsubsection{\textcolor{black}{Changes in Virtual Object Density}}
\label{subsubsec:density}
\begin{figure}[tb]
    \centering
    \includegraphics[width=.425\textwidth]{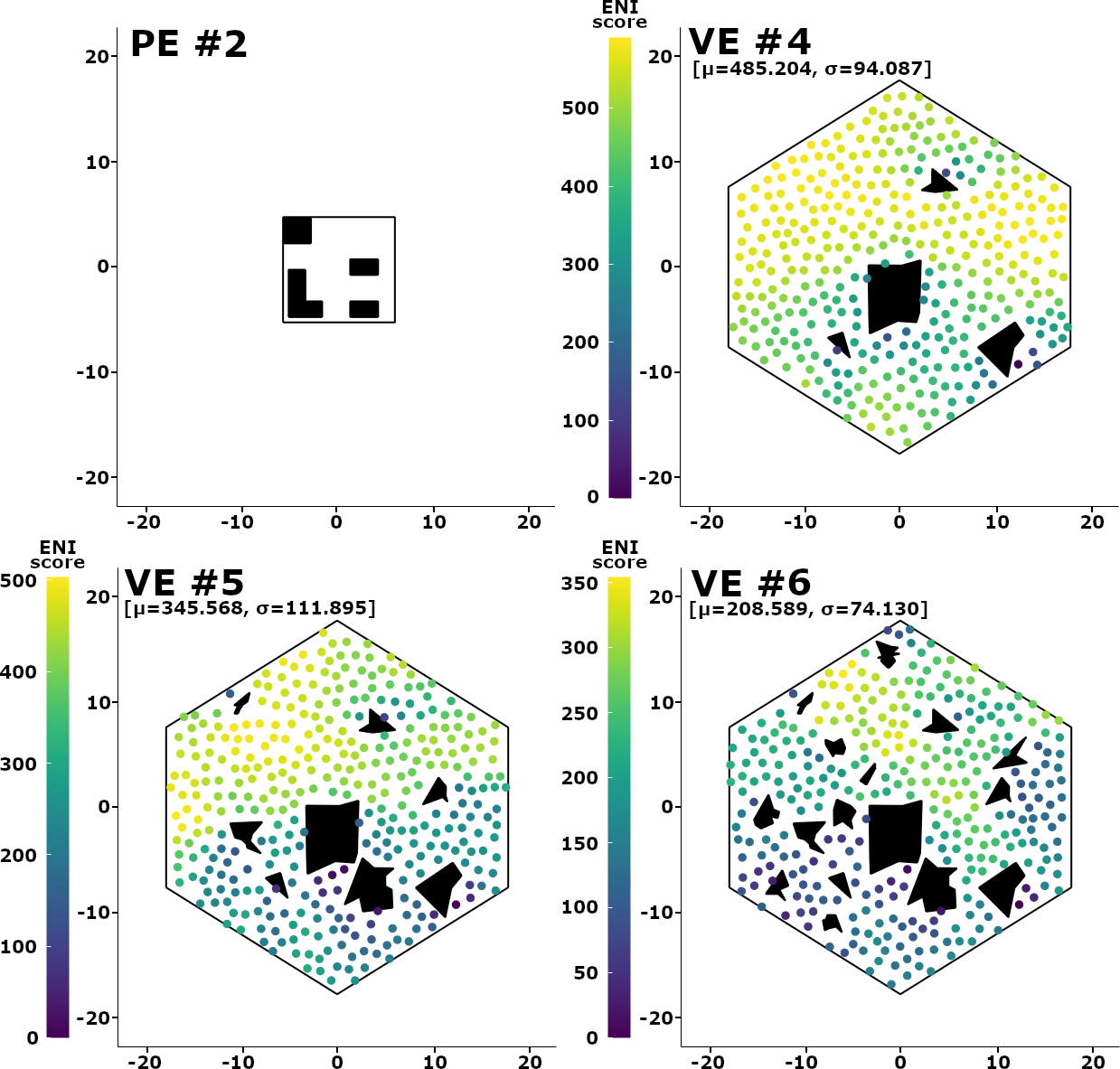}
    \caption{ENI metric scores for three different $\langle$PE,~VE$\rangle$ pairs, where the PE is static and the density of objects in the VE increases. As the density increases, the amount of navigable space decreases, creating a more navigable $\langle$PE,~VE$\rangle$ pair due to the small size of the PE.}
    \label{fig:VE_density}
\end{figure}

\textcolor{black}{
For this example, we consider the PE and VEs shown in \autoref{fig:VE_density}.
The PE is designed to represent a room that could be found in a home, such as a living room.
The VEs are chosen to have the same hexagonal boundary shape with an area of $900m^2$, but with varying amounts of random polygonal objects.
The first environment has four objects, the second has eight objects, and the third environment has sixteen objects.
Objects that are present in one VE are still present in the VEs with higher object density, to make comparisons between conditions easier.
}

\textcolor{black}{
Results of the ENI metric measurements for each of the $\langle$PE,~VE$\rangle$ pairs are shown in \autoref{fig:VE_density}.
As the VE is populated with more obstacles, the average and maximum ENI scores decrease, indicating that the more cluttered VEs are more compatible with the given PE for navigation.
Intuitively, this makes sense since the amount of space in the VE that is visible (and thus, immediately navigable) goes down as the object density increases.
As the area of immediately navigable space in the VE decreases, it approaches the area of immediately navigable space in the much smaller PE.
Results from the validation of the ENI metric (see \autoref{subsec:exp1}) confirm that navigability increases as the density of objects in the VE increases.
From this example, we can see that introducing obstacles into an environment can actually make it \textit{easier} for users to avoid collisions in the PE, if the PE also has some obstacles that may obstruct the user's path.
}

\subsubsection{\textcolor{black}{Changes in Virtual Environment Size}}
\label{subsubsec:size}
\textcolor{black}{
When investigating the effects of the size of the VE on navigability, we considered the PE and VEs shown in \autoref{fig:teaser}.
Similar to the object density experiment, we chose a fixed PE that could be found in a home (via \cite{williams2021redirected}), and constructed three VEs of varying sizes ($400 m^2, 900 m^2,$ and $1600 m^2$), each with different boundary shapes and ten different, random polygonal objects.
}

\textcolor{black}{
% The ENI metrics for each of the three $\langle$PE,~VE$\rangle$ pairs are shown in \autoref{fig}.
We see that the mean and maximum ENI values increase as the area of the VE increases.
Intuitively, this is expected since VEs with larger area are more likely to lead to longer paths that cannot be traversed from within the small PE.
% This can be understood by the fact that in environments with more open space, users are more likely to walk on longer straight-line paths which cannot fit in the small PE.
When the navigable area of the VE decreases, users are more likely to travel on virtual paths with shorter segments, since they will be forced to make more turns to avoid objects in the VE.
These turns increase the chance that the user will also turn away from nearby objects in the PE and incur fewer collisions, as the ENI scores suggest.
This result is also confirmed by the validation experiment in \autoref{subsec:exp1}, in which we estimate the navigability of the $\langle$PE,~VE$\rangle$ pairs in \autoref{fig:teaser} via simulated paths.
Thus, from this experiment, it is clear that increasing the area of the navigable space in the VE leads to more collisions in VR locomotion.
% We conducted simulations of users walking in these $\langle$PE,~VE$\rangle$ pairs to validate this conclusions, and the results for this validation can be found in \autoref{subsec:exp1}.
}

\subsection{\textcolor{black}{Design Guidelines Based on ENI}}
\textcolor{black}{
Although ENI helps us understand how the layout of the environment influence navigability of a $\langle$PE,~VE$\rangle$ pair, it is not always straightforward to translate an ENI metric into a more compatible $\langle$PE,~VE$\rangle$ pair.
In this section, we provide some high-level guidelines on how to design VEs that are more amenable for VR locomotion, relative to a given PE.
We note that these are not strict rules for designing virtual environments, but rather are intended to be suggestions on how to create environments that are more likely to yield a more comfortable navigation experience for users.
}

\textcolor{black}{
In general, collision-free locomotion in VR is most difficult when the VE contains large, open spaces while the PE is small and cluttered with objects.
This is an undesirable situation because in these cases, it is more likely that the user will travel on a long, straight path in the VE which cannot be traversed in the PE due to obstacles.
To avoid this, designers of VEs should try to place objects in the VE such that the navigable area in the VE is reduced (e.g. \autoref{subsubsec:density}).
This forces users to travel on virtual paths with more turns as they avoid virtual objects--these turns make it more likely that the user will also turn away from obstacles in the PE.
Virtual structures like narrow corridors are favorable since they restrict the number of possible paths that the user can travel along, which decreases the chance that their particular path yields a collision in the PE.
}

\textcolor{black}{
In addition to reducing the frequency of large, open spaces, designers may want to place objects in the VE that have a similar size, shape, and distribution to objects in the PE if possible.
This has the effect of making the PE and VE more similar on a local scale (\textit{i.e.} the visibility polygons are more similar), which further increases the chance that the user can travel along collision-free paths (see \autoref{fig:superimpose}).
}

%% file: validation.tex
\section{User Studies and Validation}
\label{sec:validation}
% PLEASE PUT SOME IMAGES OF DIFFERENT USERS WITHIN YOUR LAB SETUP. THAT WILL CLEARLY CONVEY HOW YOUR APPROACH WAS EVALUATED IN A REAL-WORLD SETTING
To validate the ENI metric, we collected navigation data from simulations and \textcolor{black}{two user studies}.
The goal of the ENI metric is to provide insight into the relationship between environment layouts and ease of collision-free \textcolor{black}{locomotion in VR.}
Therefore, it is expected that our metric is correlated with the navigation behavior of users or provides new insights in terms of designing \textcolor{black}{$\langle$PE,~VE$\rangle$ pairs.}
%insight that was difficult or impossible to make before.
To this end, we show that our metric correctly identifies pairs of physical and virtual configurations that allow for easier collision-free navigation (\autoref{subsec:exp1}) and that ENI is correlated with users' tendency to avoid physical objects during locomotion in VR (\autoref{subsec:exp2}).

\subsection{Experiment 1: Simulation Experiment}
\label{subsec:exp1}
% TODO:
%     - compute the similarity values (\phi(p_virt, p_phys)) for each of the 20 points I tested in each env. This will allow us to make some connection between the similarity value and the distance walked
\subsubsection{Design}
In this experiment, \textcolor{black}{we simulated a user walking along 50 paths in different pairs of physical and virtual environments.
Specifically, we tested the six environment pairs examined in \autoref{subsec:VE_changes} and the three pairs used in \cite{williams2021arc}.
In each environment, we simulated the user traveling along 50 different paths with a random start and end configuration in the VE, and a random starting configuration in the PE.
Paths through the VE were generated using the RRT* algorithm \cite{karaman2011sampling} due to its efficiency and ability to guarantee complete paths.
For each path, the simulated user would traverse the path in the VE and the PE simultaneously.
If the user got too close to an object in the PE, a reset maneuver was initiated such that they were reoriented away from the nearby object.
Users were reset with the reset-to-gradient technique presented in \cite{thomas2019general} due to its ability to work in a variety of different environments.
To quantify the navigability of the environments in accordance with the definition of navigability we adopt in this paper (see \autoref{subsec:navigability_metrics}), we computed the average distance travelled before a reset was incurred across all 50 paths in each $\langle$PE,~VE$\rangle$ pair.
}

\subsubsection{Results}
The results of the \textcolor{black}{average distance walked between resets is shown in \autoref{tab:simulation_results}.
In general, we see that as the ENI decreases (i.e. navigability improves), the average distance walked between resets increases, indicating that the simulated user was able to travel further before being interrupted by a reset.
This trend in the results confirms that ENI is correctly able to identify $\langle$PE,~VE$\rangle$ pairs that are better or worse for natural walking in VR.
}

\textcolor{black}{
Although the high-level trends showed that lower ENI is associated with increased navigability, there are some interesting results in the data.
First, we see that there is only a small difference in average distance walked between the $\langle$PE~\#1, VE \#2$\rangle$-$\langle$PE~\#1, VE \#3$\rangle$ and $\langle$PE~\#2, VE \#4$\rangle$-$\langle$PE~\#2, VE \#5$\rangle$ environment pairs, despite the large differences in their ENI scores.
While the cause of this is not clear, we believe that this plateauing effect in the distance walked might be an indication of the lower bounds on navigability for a given PE.
That is, beyond a certain ENI score, the navigability does not get significantly worse as the ENI increases because the difficulty in navigation becomes maximally constrained by the layout of the PE.
% This effect could be more pronounced due to the fact that the PE was static in these situations, suggesting that the large areas of the VE are already maximally incompatible for the given PE.
% While the cause of this is not clear, we believe that this plateauing effect in the distance walked might be an indication of the lower bounds on navigability for a given PE.
% That is, given a fixed PE, we believe it may be the case that 
}

\textcolor{black}{
Finally, we see that the $\langle$PE~\#4, VE \#8$\rangle$ pair yields the worst navigability, despite having a low ENI score.
Though initially surprising, this result makes sense when we consider the actual shape of PE~\#4 (see \autoref{fig:ARC_EnvB} in the supplementary material).
Although the PE and VE have similar local structure, the PE consists only of narrow corridors, which are inherently difficult to navigate without getting too close to any obstacles.
In this case, small deviations from a path that travels directly down a corridor are likely to lead to resets.
Additionally, the reset-to-gradient maneuver reorients users such that they face directly away from the object they got too close to.
In this particular environment, this means that the user often faces the opposite wall of the corridor, which they then walk directly towards after finishing the reset.
Thus, the simulated user gets stuck, oscillating back and forth between the walls of the corridor.
This result highlights one shortcoming of the ENI metric: by only considering the geometry of the environments, and \textit{not} considering the motion behavior of the user through the environments, some aspects of the environment structure that are important for navigability cannot be properly evaluated using our metric.
}

\setlength{\tabcolsep}{5pt}
\begin{table}[ht]
    \centering
    \begin{tabular}{lllll}
         & \multicolumn{2}{c}{\textbf{Distance Between Resets}} & \multicolumn{2}{c}{\textbf{ENI score}} \\ \midrule \midrule
        $\langle$PE,~VE$\rangle$ pair & $\mu$ ($m$) & $\sigma$ ($m$) & $\mu$ & $\sigma$ \\ \midrule
        $\langle$PE~\#1, VE \#1$\rangle$& 3.337 & 1.869 & 80.297 & 32.020 \\ % 20x20 (size scale)
    $\langle$PE~\#1, VE \#2$\rangle$& 3.082 & 1.802 & 241.510 & 56.741 \\ % 30x30 (size scale)
    $\langle$PE~\#1, VE \#3$\rangle$& 2.991 & 1.701 & 760.146 & 245.238 \\ % 40x40 (size scale)
        \hline
    $\langle$PE~\#2, VE \#4$\rangle$& 3.031 & 2.382 & 485.231 & 96.049 \\ % 4 objs (density scale)
    $\langle$PE~\#2, VE \#5$\rangle$& 3.065 & 2.515 & 345.408 & 112.748 \\ % 8 objs (density scale)
    $\langle$PE~\#2, VE \#6$\rangle$& 3.505 & 2.347 & 206.276 & 75.778 \\ % 16 objs (density scale)
        \hline
    $\langle$PE~\#3, VE \#7$\rangle$& 6.075 & 2.835 & 0.530 & 0.251 \\ % Env A (ARC)
    $\langle$PE~\#4, VE \#8$\rangle$& 0.989 & 0.878 & 9.101 & 4.836 \\ % Env B (ARC)
    $\langle$PE~\#5, VE \#9$\rangle$& 3.129 & 1.993 & 78.357 & 33.292 \\ % Env C (ARC)
        \midrule \midrule
    \end{tabular}
    \caption{\textcolor{black}{Navigability results from simulating 50 random walking paths in different $\langle$PE,~VE$\rangle$ pairs. Here, we define navigability as the average distance that the user can walk in the VE before colliding with an object in the PE, across all configurations in the PE and VE (\autoref{subsec:navigability_metrics}). In general, the navigability of the environments decreases as the ENI score increases, indicating that our metric is able to correctly identify $\langle$PE,~VE$\rangle$ pairs that are less amenable to real walking.}
    \label{tab:simulation_results}}
\end{table}

\subsection{Experiment 2: User Studies}
\label{subsec:exp2}
% YOU SHOULD ORGANIZE THIS SECTION AS:
% (i) WHAT IS THE UNDERLYING HYPOTHESIS THAT YOU WANTED TO EVALUATE?
% (ii) WHAT ARE THE ISSUES IN TERMS OF DESIGN OF THE ENVIRONMENT? HOW DO YOU SELECT THE ENVIRONMENTS, USER'S PATH AND OTHER THINGS THAT HELP IN TERMS OF EVALUATING THE METRIC EFFECTIVENESS.
In our second experiment, we hypothesized that $\langle$PE,~VE$\rangle$ pairs with higher ENI scores will cause users to incur more resets than $\langle$PE,~VE$\rangle$ pairs with low ENI scores.

\subsubsection{Design}
\begin{figure}[t]
    \centering
    \includegraphics[width=.42\textwidth]{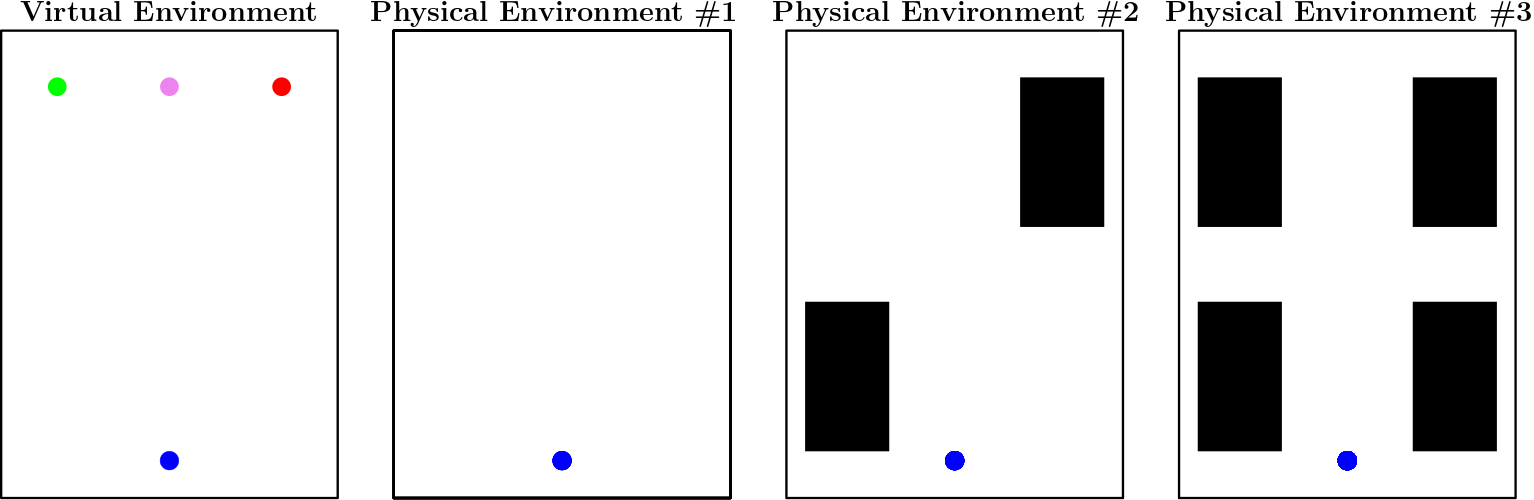}
    \caption{Diagrams of the layouts of the VE and PEs used in the first user study. The blue circle indicates the user's starting position in each environment, and the green, pink, and red circles indicate the locations of the goal in the VE during different trials. The dimensions of each environment are $4.37m \times 6.125m$.}
    \label{fig:user_study_envs}
\end{figure}

\textcolor{black}{
We conducted two user studies.
In both studies, participants were tasked with reaching a goal in the VE, indicated by a floating yellow block.
In the first study, participants navigated through a VE with no obstacles, while the layout of the PE changed to have increasing density of objects.
Participants completed the walking task a total of three times for each $\langle$PE,~VE$\rangle$ pair, with the goal location being different in each of the three trials (see \autoref{fig:user_study_envs}).
The dimensions of the PE and VE were the same ($4.37m \times 6.125m$).
Twenty people participated in the first study (8 female (age $\mu=23,\sigma=2.6$), 11 male (age $\mu=23.4,\sigma=2.5$), 1 non-binary (age 25)).
}

\begin{figure}[t]
    \centering
    \includegraphics[width=.4\textwidth]{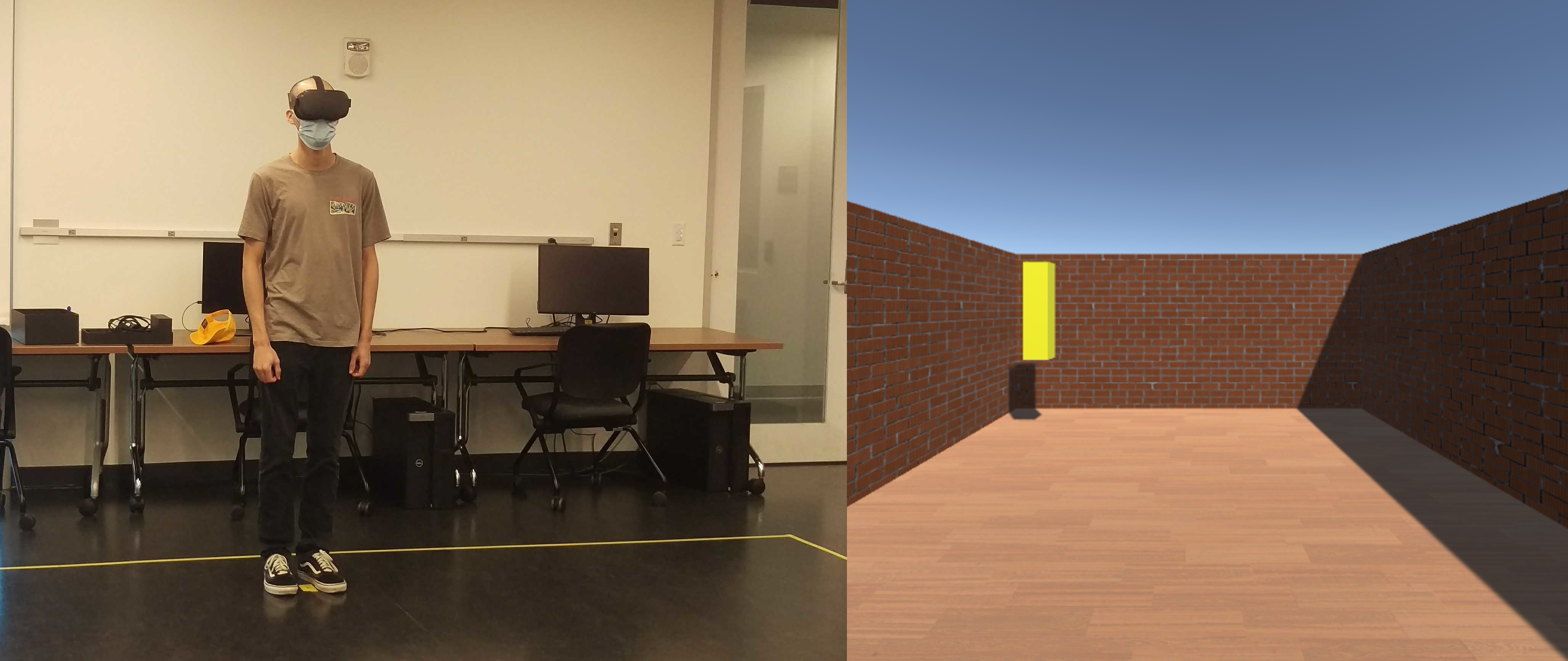}
    \caption{\textit{Left:} A user in the lab space in which we conducted our user evaluations. \textit{Right:} A screenshot of the user's starting configuration in the VE at the beginning of a trial in our \textcolor{black}{first} user study.}
    \label{fig:user_study_lab}
\end{figure}

\textcolor{black}{
In the second study, participants navigated through three different VEs, each with a varying size (similar to \autoref{fig:teaser}).
The PE was the same for each of the three VEs.
Participants completed a total of three walking trials, each experiencing a random goal location in each trial (the same for each participant).
The order in which participants experienced the $\langle$PE,~VE$\rangle$ pairs was the same across participants.
A total of 10 people participated in the second study (9 male (age $\mu=25, \sigma=3.5$), 1 female (age 25)).
Full details on the studies and diagrams of the $\langle$PE,~VE$\rangle$ pairs used in the two experiments can be found in the supplementary materials.
}

\subsubsection{Results}
\textcolor{black}{
The results of the two user studies can be seen in \autoref{tab:user_study_results}.
We validate our metric by measuring the navigability (average distance walked between resets) as the ENI changes.
Results show that as the ENI increases (i.e. $\langle$PE,~VE$\rangle$ compatibility decreases), users walked less distance between resets (i.e. navigability decreased).
This confirms that our ENI metric is correctly able to identify $\langle$PE,~VE$\rangle$ pairs which have comparatively more or less navigability.
}

\textcolor{black}{
Interestingly, the trends in the results from the user studies do not show the same plateauing effect as the results from \autoref{subsec:exp1}.
In the simulation experiments, our results show that two $\langle$PE,~VE$\rangle$ pairs can have very large differences in ENI scores, but can yield very similar distances walked between resets.
While we are not certain why the plateauing effect did not appear in the user study results, we believe it may be due to the lower total number of paths collected, or due to the smaller differences in ENI scores between $\langle$PE,~VE$\rangle$ pairs, compared to the differences in the simulation experiments.
Future work should study this plateauing effect in more detail.
}

\setlength{\tabcolsep}{5pt}
\begin{table}[ht]
    \centering
    \begin{tabular}{lllll}
         & \multicolumn{2}{c}{\textbf{Distance Between Resets}} & \multicolumn{2}{c}{\textbf{ENI score}} \\ \midrule \midrule
        $\langle$PE,~VE$\rangle$ pair & $\mu$ ($m$) & $\sigma$ ($m$) & $\mu$ & $\sigma$ \\ \midrule
        $\langle$PE~\#6, VE \#10$\rangle$ & 3.217 & 1.398 & 0.530 & 0.251 \\ % Study 1 Env A (empty PE)
        $\langle$PE~\#7, VE \#10$\rangle$ & 2.927 & 1.548 & 8.072 & 1.013 \\ % Study 1 Env B (2 obs PE)
        $\langle$PE~\#8, VE \#10$\rangle$ & 2.390 & 1.153 & 14.097 & 0.612 \\ % Study 1 Env C (4 obs PE)
        \hline
        $\langle$PE~\#9, VE \#11$\rangle$ & 4.098 & 2.168 & 27.131 & 8.864 \\ % Study 2 Env A (10x10 VE)
        $\langle$PE~\#9, VE \#12$\rangle$ & 3.688 & 2.604 & 102.129 & 24.390 \\ % Study 2 Env B (15x15 VE)
        $\langle$PE~\#9, VE \#13$\rangle$ & 3.559 & 2.302 & 217.361 & 48.958 \\ % Study 2 Env C (20x20 VE)
        \midrule \midrule
    \end{tabular}
    \caption{\textcolor{black}{Navigability results from two separate user studies. In the first user study (first three rows), users walked towards a goal location in a static VE while located in three different PEs. In the second study (bottom three rows), users searched for a goal location in different VEs while located in the same PE. In both situations, our results showed that navigability decreases as the ENI score increases, validating the correctness of our metric.}
    \label{tab:user_study_results}}
\end{table}

%% file: conclusion.tex
\section{Conclusion, Limitations, \& Future Work}
\label{sec:conclusion}
In this work, we presented Environment Navigation Incompatibility (ENI) metric, a novel metric for quantifying the navigability of a pair of physical and virtual environments\textcolor{black}{, based on their geometric layouts.}
ENI measures the navigability of a $\langle$PE,~VE$\rangle$ pair by measuring the similarity (compatibility) of the two environments, since collisions during locomotion in VR are mainly caused by differences in the layouts of the PE and VE.
% Our metric is based on the geometric layouts of the environments.
%meaning it avoids ambiguity that can arise from being defined on properties that do not fully characterize the layout of an environment, such as area.
By uniformly sampling the environments and computing visibility polygons at sampled points, our metric accurately captures the features of the environments that are amenable to collision-free navigation (namely, the local surroundings of a user across the PE and VE).
We validate our metric through simulations and \textcolor{black}{two user studies}, showing that ENI can accurately identify $\langle$PE,~VE$\rangle$ pairs that are more amenable to collision-free navigation \textit{without} requiring locomotion data. 
In general, users \textcolor{black}{were able to walk further before incurring a reset in environment pairs that our metric identified as more navigable.}
 
Although the ENI metric is effective at identifying compatible $\langle$PE,~VE$\rangle$ pairs for RDW, it has some  limitations.
First, the computation time can be long if the environments \textcolor{black}{have a large number of sampled points} or have a high number of obstacles.
\textcolor{black}{Our implementation was done in Python and was not parallelized, so there is room for significant speed-up by porting the implementation to a faster language and by taking advantage of multithreading for visibility polygon computation.}
Another limitation of ENI is that it is currently limited to static environments.
Extensions of ENI to environments with dynamic obstacles will likely require adding a temporal component to the computation, which may increase the computation time even further.
Additionally, ENI provides a ``best case'' mapping of virtual configurations to physical configurations. This best case mapping can be misleading for assessing navigability in some cases, since many virtual configurations can map onto the same physical configuration.
% This is a problem since it may not be possible to apply RDW to steer the user to the most compatible physical configuration if it is too close to the current physical configuration, as can be the case when multiple virtual configurations map to the same physical configuration.
Another limitation of ENI is that it only considers the layouts of the environments, and not any other factors that are known to influence the navigation experience during \textcolor{black}{locomotion}, such as the \textcolor{black}{specific} paths travelled.
% and the user's subjective perceptual thresholds.
Finally, the validation experiments were limited in that we could \textcolor{black}{not test ENI on a large corpus of environment pairs.}
Although we believe this does not affect the validity of our metric, it is important to evaluate the accuracy of any metric in as many scenarios as possible.

There are many avenues for future work in this area.
Since our experiments showed mixed results in terms of the correlation between ENI scores and navigability measures, we would like to further study the ENI metric with a larger set of benchmark environments to get a better understanding of the relationship between our ENI metric and the navigability of environment pairs.
Additionally, more user studies should be conducted to investigate whether or not ENI aligns with users' subjective perception of the navigability, \textcolor{black}{though this will require careful consideration to ensure that all participants have the same notion of ``navigability.''
Another area for future work is to investigate the cause of the plateauing effect we saw in navigability scores in \autoref{subsec:exp1}.
A detailed understanding of the worst-case navigability for given $\langle$PE,~VE$\rangle$ pairs may allow us create standardized benchmarks against which we can compare the efficacy of different locomotion interfaces.
}
Finally, if the computation time for ENI can be improved to interactive rates, we believe that it will also be interesting to evaluate the effectiveness of using ENI to help optimize the layouts of VEs to make them more amenable to navigation (similar to other architecture tools like Goldstein et al. \cite{goldstein2020spaceanalysis}).

% - computation takes quite a while for high sample densities
% - this provides "best case" mapping: multiple virt positions can map to the same physical location. this is a problem when the user walks along a path (As is normal) in the VE, since the "best phys pos" for each pos along the virtual path may map t oteh same phys pos. can mke it impossible to steer the user along the best phys pos for each virt pos of the path using RDW (because RDW is not strong enough)
% - only works in static scenes
% - no modelling of the human, which does end up having a large influence on the locomotion experience
% - sample size in this paper is kinda small. there are an infinite number of envs out there, so future research should work on testing out our metric in other types of environments
%     - simiarly, the user study was kinda small (only 3 PEs and 1 VE)

% \textbf{todo: talk about how the interactive visualization opens the door for creating a sort of "rdw environment designer" assissted tool to make things easier for creators.}

%% file: supplementary.tex
\newpage 
\newpage
\clearpage

\renewcommand{\thesection}{\Alph{section}}
\setcounter{section}{0}
\section{Supplementary Materials} 
\label{sec:supplementary}

\subsection{Additional Applications of ENI}
\label{subsec:supp_additional_applications}
Here, we further demonstrate the usefulness of ENI by showing how it can be used to analyze the performance of redirected walking (RDW) controllers.

\subsubsection{Performance Analysis of RDW Controllers}
% - rdw controllers try to steer the user along collision-free paths
% - typical performance metrics for these controllers are the number of resets incurred or the virtual distance walked between resets.
% - we can use our compatibility metric to get a measure of "average compatibility" of the physical and virtual paths that a user is steered along with a given RDW controller
% - compute the vis poly at discrete poitns along the phys/virt paths. compute the compatibility score for each pair of phys/virt vis polys along the path. this gives us avg compatibility
% - opens the door for developing new RDW controllers that try to optimize for compatibility rather than minimal number of resets. this could be preferred because it is often difficult to determine what is the actual cause of resets during walking, since they happen somewhat infrequently. compatibility metric, on the other hand, can provide a measure of performance goodness at a faster rate, so you get more real-time feedback on performance.
Redirection controllers are algorithms that are used to compute the correct redirection that is applied to optimally steer the user away from physical objects~\cite{nilsson201815}.
% When the user gets too close to a physical object, a reset maneuver is initiated to reorient the user away from the object before they start to walk again.
% Thus, a RDW controller aims to minimize the number of resets a user incurs.
The performance of controllers is typically evaluated by measuring the number of times the user is reset after getting too close to a physical object \cite{williams2021redirected, williams2021arc, thomas2019general, messinger2019effects,bachmann2019multi}, the average distance the user is able to walk before initiating a reset \cite{strauss2020steering,thomas2019general,williams2021arc,bachmann2019multi}, or the average strength of redirection during locomotion \cite{lee2020optimal,williams2021arc,bachmann2019multi}.
% GIVE REFERENCES ON SOME CONTROLLERS WHICH DISCUSS THESE RESET BEHAVIORS.
These performance metrics are useful, but they do not provide any information in terms of how similar is the user's proximity to obstacles in the PE and VE.
By calculating the ENI measure between corresponding physical and virtual configurations along a given path through both environments, we can gain an understanding of the differences in the user's physical and virtual proximity to objects.
This kind of performance metric provides information about the user's locomotion experience (for a given RDW controller) that cannot be derived from traditional performance metrics such as number of resets or intensity of redirection.

\begin{figure}[htb]
    \centering
    \includegraphics[width=.45\textwidth]{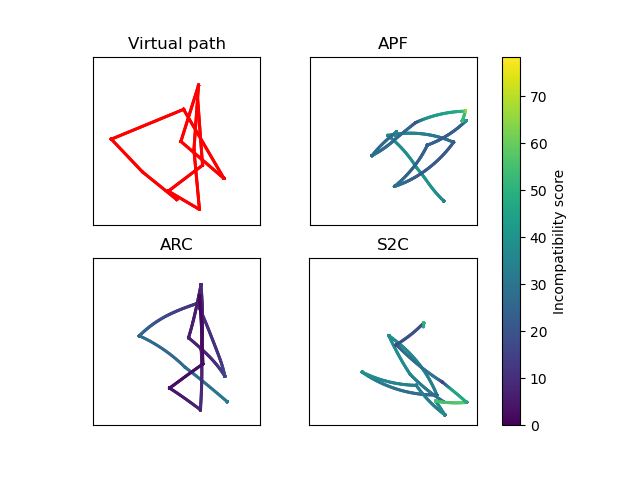}
    \caption{\textit{Top left:} A virtual path in an empty $10m \times 10m$ VE. \textit{Top right, bottom left, and bottom right:} The physical path the user travels on when steered by APF \cite{thomas2019general}, ARC \cite{williams2021arc}, and S2C \cite{hodgson2008redirected}. The physical paths are colored according to the ENI scores between the corresponding points along the physical and virtual paths. The path yielded from ARC is more compatible with the virtual path, suggesting that the user is less likely to incur collisions during locomotion with ARC.}
    \label{fig:path_compat}
\end{figure}

In \autoref{fig:path_compat}, we show an example result of computing the ENI of a user's configurations across a path.
In this example, the user is located in an identical $10m \times 10m$ $\langle$PE,~VE$\rangle$ pair with no obstacles.
We simulated the user walking on the same virtual path while being steered with three different RDW controllers: artificial potential fields (APF) \cite{thomas2019general}, alignment-based redirection controller (ARC) \cite{williams2021arc}, and steer-to-center (S2C) \cite{hodgson2008redirected}.
For the path segment shown in \autoref{fig:path_compat}, the ENI measures were $[\mu = 34.387, \sigma = 11.680]$ for APF, $[\mu = 11.468, \sigma = 8.513]$ for ARC, and $[\mu = 34.177, \sigma = 9.532]$ for S2C.
This result suggest that when steered by ARC, the user is, on average, in a physical configuration that is more compatible with their virtual configuration than when they are steered with APF or S2C.
That is, for most configurations along the path, the user is more likely to be able to travel on a collision-free path when steered by ARC than by APF or S2C. 
In the short path segment we considered in this example, this is indeed the case as the user incurs zero collisions when steered by ARC, but incurs two resets when steered by APF and S2C.
ARC was designed to steer the user in an attempt to match their physical and virtual proximity~\cite{williams2021arc}, so our metric's performance matches the expected behavior of ARC.
%so we see a this result is not surprising.
This conclusion is further supported by comparing the shape of the virtual path to each of the physical paths.
It is clear from \autoref{fig:path_compat} that the user's physical path when steered by ARC is more similar to the virtual path than the paths generated by APF and S2C algorithms.
% From this example, we can see how the ENI measure can be applied to walked paths to provide another metric with which we can evaluate the performance of RDW controllers.

\subsection{User Study Details}
During our user studies, if the user got within $0.25m$ of an obstacle, a reset was initiated and they were instructed to turn $180^\circ$ in the PE, while the virtual camera did not rotate.
For the first study, participants completed the task in blocks organized by PE (all three tasks in one PE were completed before changing to the next PE), and the order of PEs was counterbalanced.
In the second study, participants completed the navigation task once in three different VEs, in the same order across all participants.
To increase the variety in the travelled paths, each trial started with the user facing a random direction in the VE (participants all experienced the same random direction for a particular trial, but the directions were random from trial to trial).
Participants also completed a practice trial at the start of the experiment to get them accustomed to the hardware and experiment task.

A total of twenty participants successfully completed the first experiment (8 female (age $\mu=23,\sigma=2.6$), 11 male (age $\mu=23.4,\sigma=2.5$), 1 non-binary (age 25)).
The first study took about $20$ minutes for each participant, and they were compensated with a $\$10$ Amazon gift card.
After completion of the experiment, all participants completed the Kennedy-Lane Simulator Sickness Questionnaire (SSQ) \cite{kennedy1993simulator}; the largest reported SSQ score was 33.66 ($\mu=8.727,\sigma=9.494$).
Ten participants completed the second experiment (9 male (age $\mu=25, \sigma=3.5$), 1 female (age 25)).
The second study took about $15$ minutes for each participant, and they were compensated with a $\$10$ Amazon gift card.
After completion of the second experiment, all participants completed the Kennedy-Lane Simulator Sickness Questionnaire (SSQ) \cite{kennedy1993simulator}; the largest reported SSQ score was 29.92 ($\mu=7.106,\sigma=10.043$).
Both studies were approved by the authors' university's Institutional Review Board.

\newpage
\clearpage
\subsection{Additional Figures}

\begin{figure}[htb]
    \centering
    \includegraphics[width=.9\linewidth]{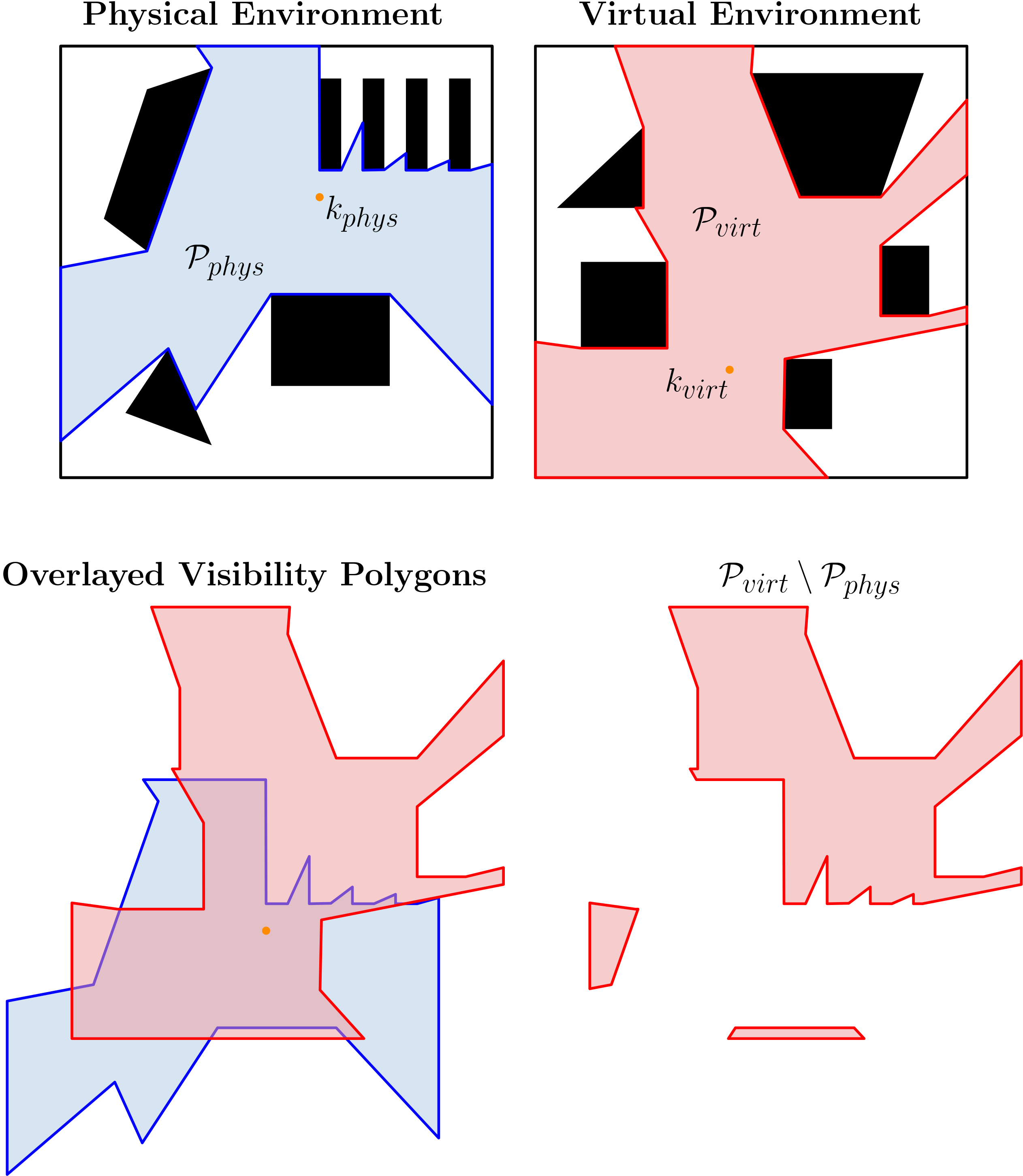}
    \caption{\textit{Top row:} Two visibility polygons $\mathcal{P}_{phys}$ and $\mathcal{P}_{virt}$ in a $\langle$PE,~VE$\rangle$ pair. \textit{Bottom row (left):} $\mathcal{P}_{phys}$ and $\mathcal{P}_{virt}$ have been translated such that their kernels lie on the same 2D position in the plane. \textit{Bottom row (right):} The result of the boolean difference operation $\mathcal{P}_{virt} \setminus \mathcal{P}_{phys}$ is shown as the red polygons. These polygons represent all the regions of $\mathcal{P}_{virt}$ that cannot be accessed when the user is located at $k_{phys}$ and $k_{virt}$ in the PE and VE, respectively. 
    Our metric uses the total area of $\mathcal{P}_{virt} \setminus \mathcal{P}_{phys}$ as a measure of the similarity of the user's local physical and virtual surroundings.
    }
    \label{fig:boolean_difference}
\end{figure}

\onecolumn
\begin{figure*}[b]
    \centering
    \includegraphics[width=.95\textwidth]{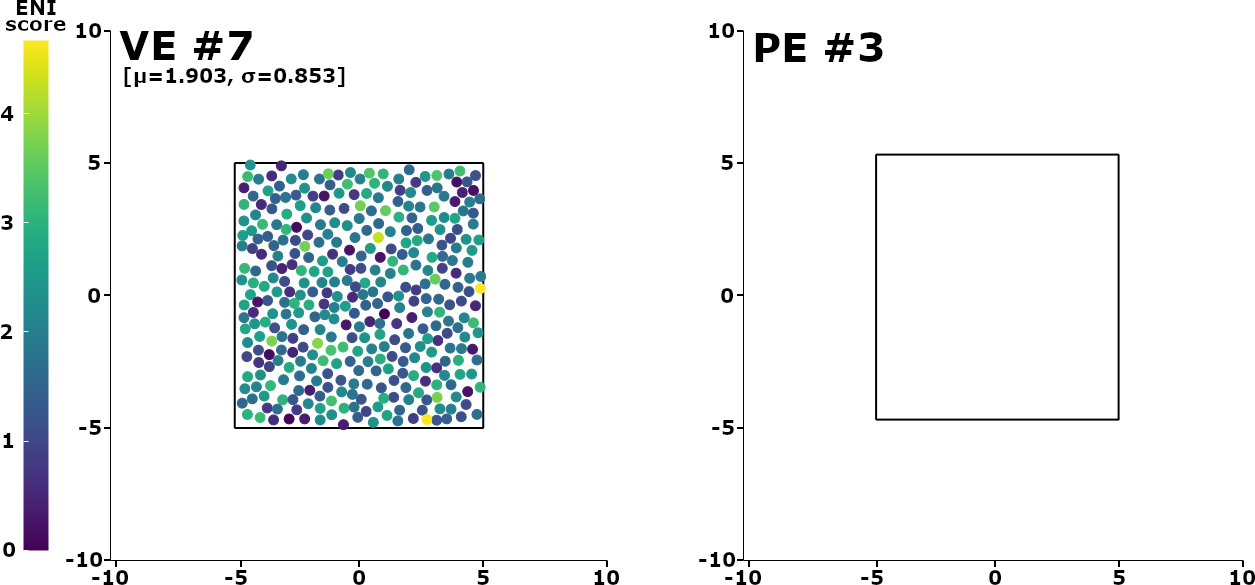}
    \caption{Environment A introduced by Williams et al. in \cite{williams2021arc}.}
    \label{fig:ARC_EnvA}
\end{figure*}

\begin{figure*}[t]
    \centering
    \includegraphics[width=.95\textwidth]{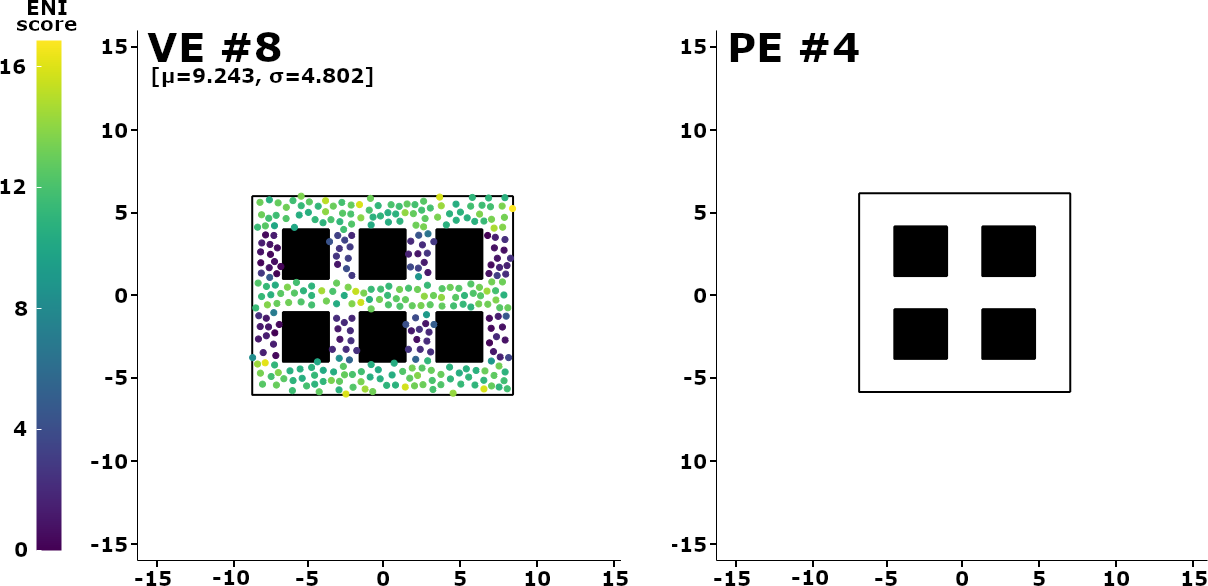}
    \caption{Environment B introduced by Williams et al. in \cite{williams2021arc}.}
    \label{fig:ARC_EnvB}
\end{figure*}

\begin{figure*}[t]
    \centering
    \includegraphics[width=.95\textwidth]{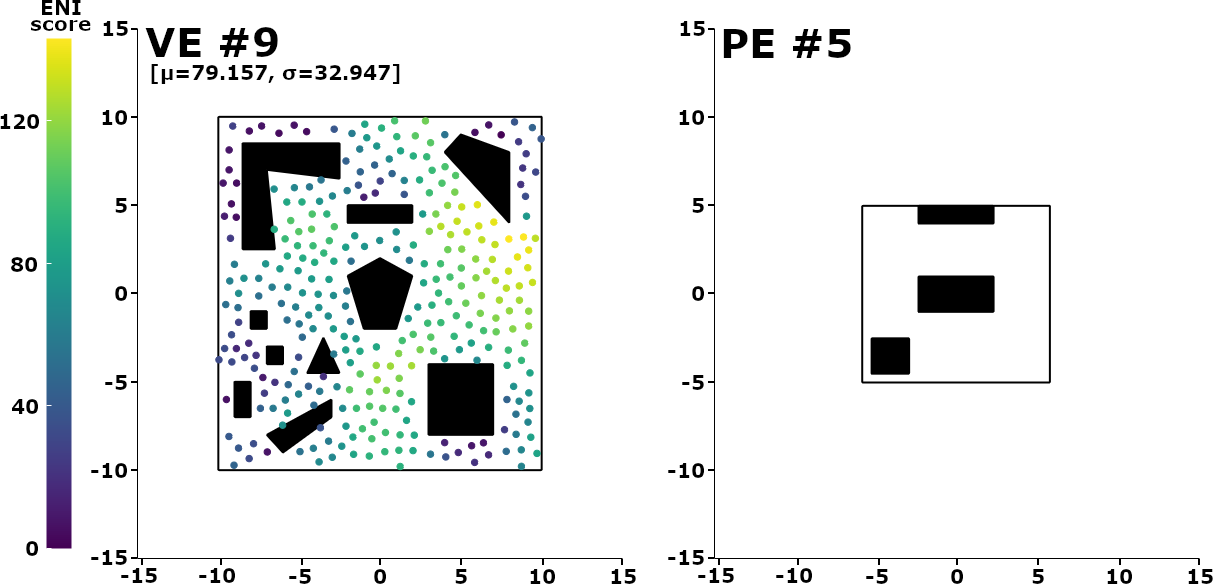}
    \caption{Environment C introduced by Williams et al. in \cite{williams2021arc}.}
    \label{fig:ARC_EnvC}
\end{figure*}

\begin{figure*}[t]
    \centering
    \includegraphics[width=.95\textwidth]{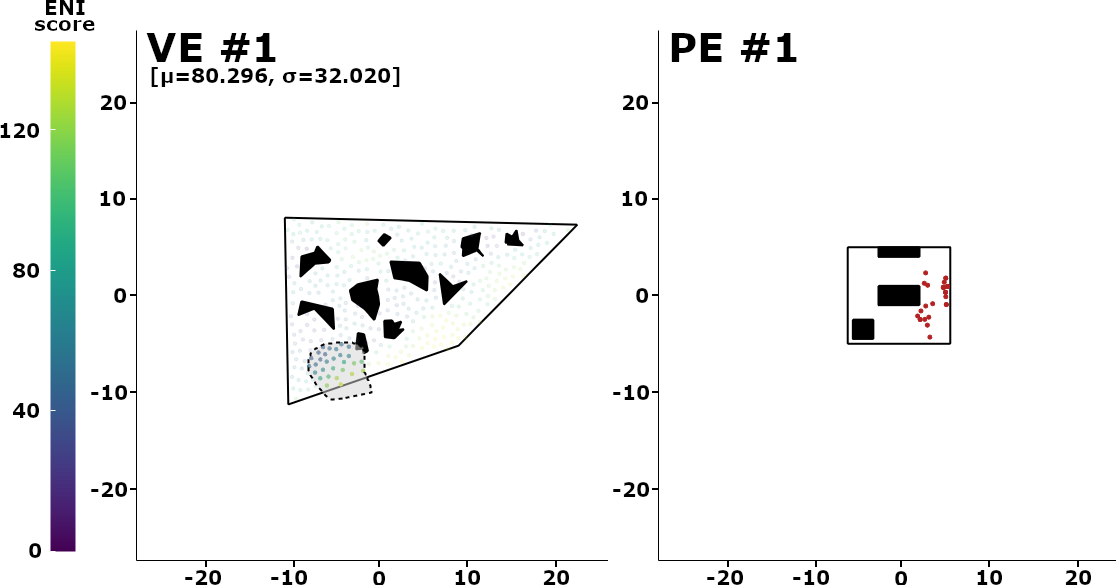}
    \caption{The effect of selecting a set of virtual points using the lasso tool. When virtual points are selected (left), the corresponding most compatible points (computed via \autoref{eqn:final_optimization}) are shown in red in the PE (right).}
    \label{fig:viz_lasso}
\end{figure*}

\begin{figure*}[t]
    \centering
    \includegraphics[width=.95\textwidth]{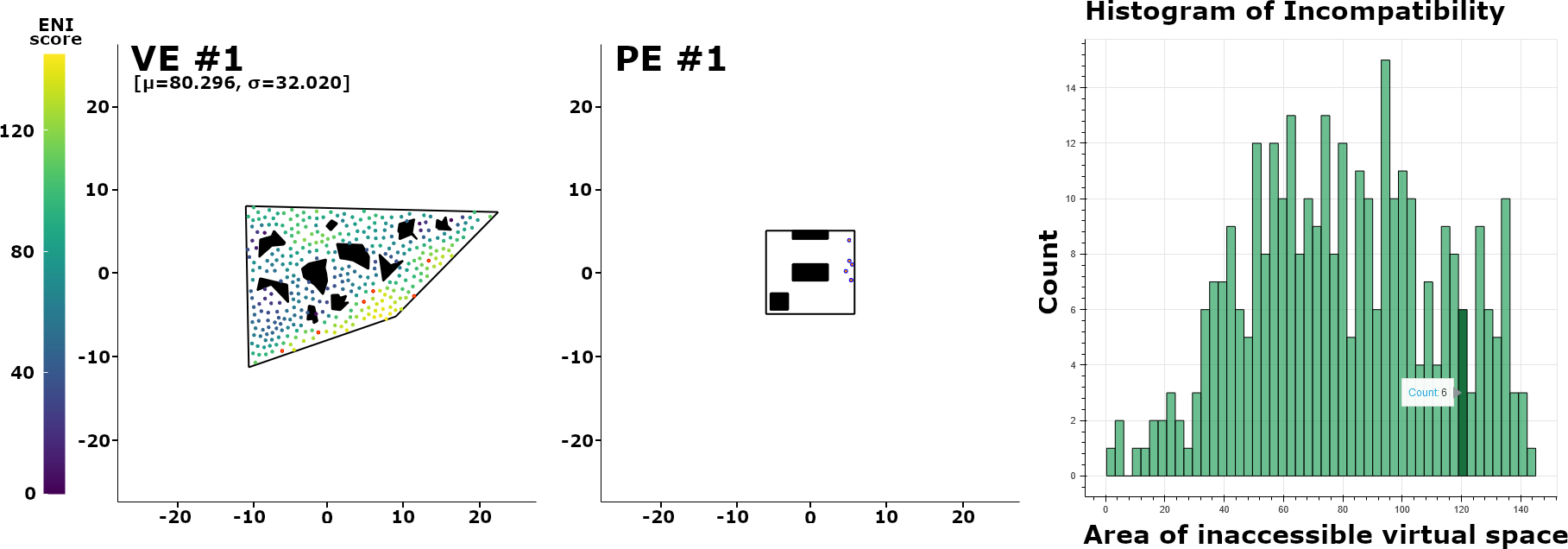}
    \caption{The effect of selecting a bar of the histogram in our interactive visualization. When a bar is selected (right), the physical and virtual points that contribute towards this histogram bar are highlighted in orange (left and middle).}
    \label{fig:vis_histogram}
\end{figure*}

\begin{figure}[htb]
    \centering
    \includegraphics[width=.9\textwidth]{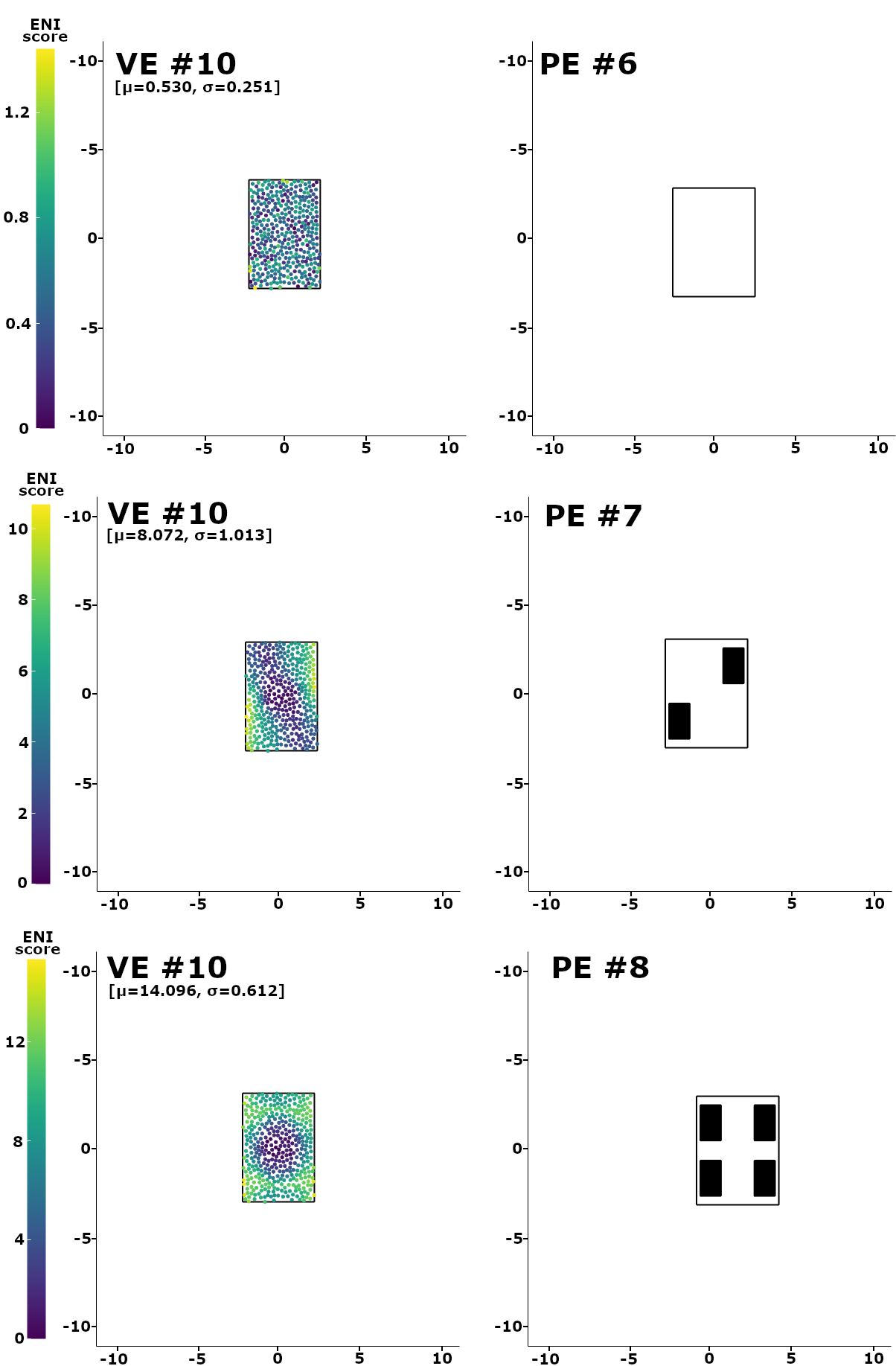}
    \caption{The three environment pairs used in our first user study.}
    \label{fig:study1_envs}
\end{figure}

\begin{figure*}[htb]
    \centering
    \includegraphics[width=.9\textwidth]{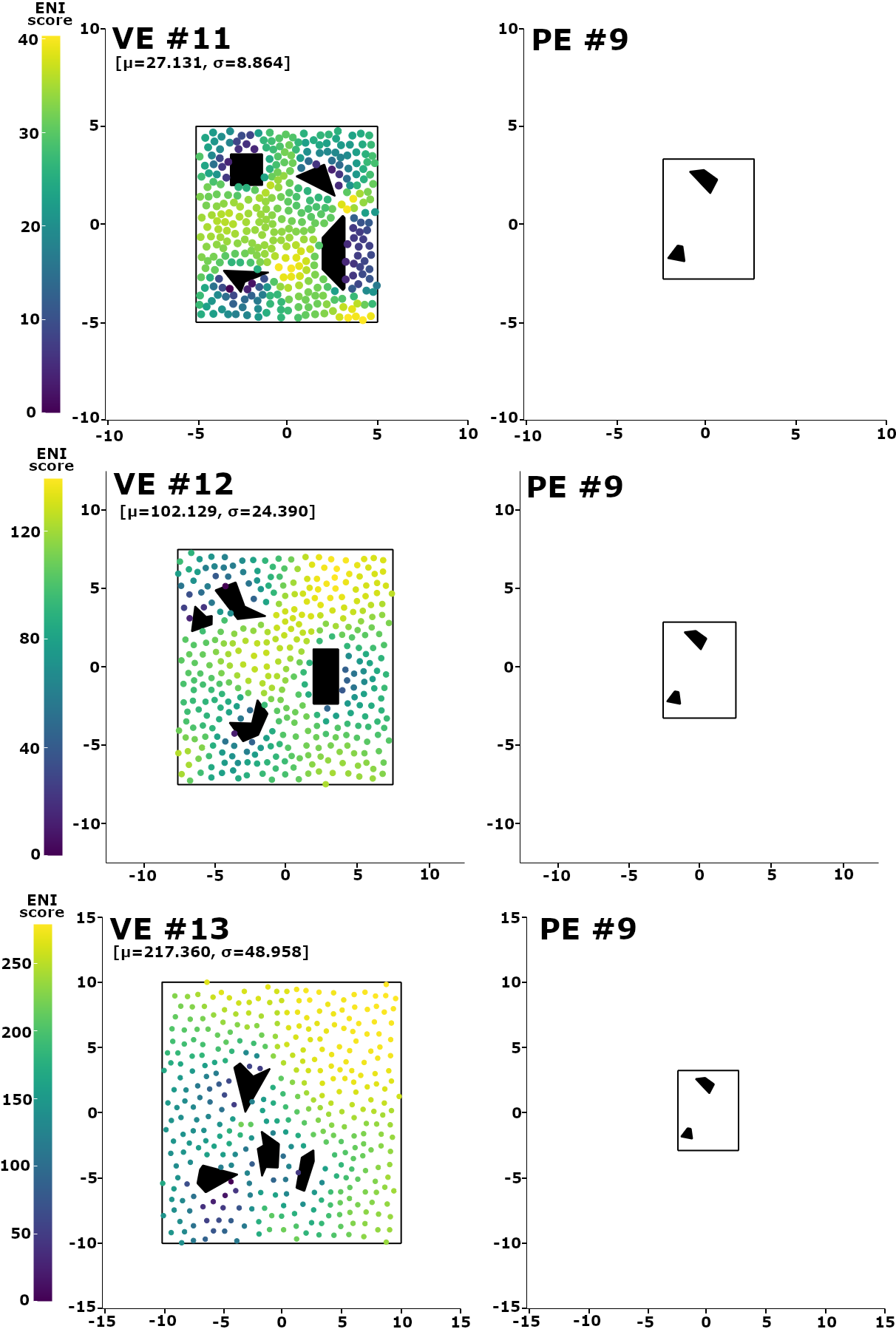}
    \caption{The three environment pairs used in our second user study.}
    \label{fig:study1_envs}
\end{figure*}